\documentclass[11pt,a4paper]{article}
\usepackage[colorlinks=true, linkcolor=black!50!blue, urlcolor=blue, citecolor=blue, anchorcolor=blue]{hyperref}
\usepackage[font=small,labelfont=bf,margin=0mm,labelsep=period,tableposition=top]{caption}
\usepackage[a4paper,top=3cm,bottom=2.5cm,left=2.5cm,right=2.5cm,bindingoffset=0mm]{geometry}

\usepackage{graphicx,placeins}
\usepackage{float}
\usepackage{afterpage}
\usepackage{epsfig,cite}
\usepackage{amssymb}
\usepackage{amsmath}
\usepackage{dsfont}
\usepackage{multirow}
\usepackage{url}
\usepackage{xcolor,colortbl}
\usepackage{float}
\usepackage{afterpage}
\usepackage{url}
\usepackage{hyperref}
\usepackage{booktabs}
\usepackage{mathrsfs}

\usepackage{tikz}
\usepackage{tikz-3dplot}
\usepackage[compat=1.0.0]{tikz-feynman}

\usepackage{enumitem}
\usepackage{hyperref}
\usepackage{cite}

\usepackage{pifont}

\usetikzlibrary{shapes, arrows}
\usetikzlibrary{decorations.pathreplacing}
\usetikzlibrary{positioning, calc}
\tikzstyle{fitted} = [rectangle, minimum width=5cm, minimum height=1cm, text centered, draw=black, fill=red!30]
\tikzstyle{operations} = [rectangle, rounded corners, minimum width=2cm,text centered, draw=black, fill=red!30]
\tikzstyle{roundtext} = [rectangle, rounded corners, minimum width=2cm, minimum height=0.8cm, text centered, draw=black, fill=red!30]
\tikzstyle{n3py} = [rectangle, rounded corners, minimum width=3cm, minimum height=1cm, text centered, draw=black, fill=green!30]
\tikzstyle{myarrow} = [thick,->,>=stealth]
\tikzstyle{line} =[draw, -latex']
\tikzstyle{decision} = [diamond, draw, fill=red!20, text width=7.5em, text centered,  inner sep=0pt, minimum height=2em, aspect=4]
\tikzstyle{cloud} = [draw, ellipse,fill=green!20, minimum height=2em]
\tikzstyle{inout} = [rectangle, draw, fill=green!20, text width=9.5em, text centered, rounded corners, minimum height=2em, minimum width=10em]
\tikzstyle{block}=[rectangle, draw, fill=blue!20, text width=9.5em, 
                   text centered, rounded corners, minimum height=2em, 
                   minimum width=10em]

\definecolor{darkgreen}{rgb}{0.0, 0.5, 0.13}

\newcommand{\be}{\begin{equation}}
\newcommand{\ee}{\end{equation}}
\newcommand{\bea}{\begin{eqnarray}}
\newcommand{\eea}{\end{eqnarray}}
\newcommand{\bi}{\begin{itemize}}
\newcommand{\ei}{\end{itemize}}
\newcommand{\ben}{\begin{enumerate}}
\newcommand{\een}{\end{enumerate}}

\def\gsim{\mathrel{\rlap{\lower4pt\hbox{\hskip1pt$\sim$}}
    \raise1pt\hbox{$>$}}}         
\def\lsim{\mathrel{\rlap{\lower4pt\hbox{\hskip1pt$\sim$}}
    \raise1pt\hbox{$<$}}}         

\newcommand{\draft}[1]{}

\def\beq{\begin{equation}}
\def\eeq{\end{equation}}

\def\lapprox{\lower .7ex\hbox{$\;\stackrel{\textstyle <}{\sim}\;$}}
\def\gapprox{\lower .7ex\hbox{$\;\stackrel{\textstyle >}{\sim}\;$}}

\usepackage{tabularx}
\newcolumntype{C}[1]{>{\centering\arraybackslash}p{#1}}

\usepackage{amsmath}
\usepackage{amsfonts}
\usepackage{amssymb}
\usepackage{dsfont}
\usepackage{pifont}
\usepackage{booktabs}
\usepackage{graphicx}
\usepackage{epstopdf}
\usepackage{epsfig}
\usepackage{framed}
\usepackage{makeidx}
\usepackage{siunitx}
\usepackage[capitalise]{cleveref}
\usepackage{hyperref}
\usepackage{placeins}
\usepackage[font=small,labelfont=bf]{caption}

\RequirePackage{csquotes}
\usepackage{bbm}

\begin{document}
\newgeometry{top=1.5cm,bottom=1.5cm,left=1.5cm,right=1.5cm,bindingoffset=0mm}

\vspace{2.3cm}
\begin{flushright}

 DESY-24-134\\TIF-UNIMI-2024-17\\Edinburgh 2024/9\\CERN-TH-2024-167
\end{flushright}

\begin{center}

  {\Large \bf Combination of aN$^3$LO PDFs\\[0.23cm] and implications for Higgs
    production cross-sections at the LHC}
  \vspace{1.1cm}
  
   {\bf The MSHT Collaboration}: \\[0.1cm]
   Thomas Cridge$^1$,
   Lucian A.\ Harland-Lang$^2$,
   Jamie McGowan$^2$, and
   Robert S.\ Thorne$^2$
   
    \vspace{5pt}

    \vspace{1cm}

   {\bf The NNPDF Collaboration}: \\[0.1cm]

   Richard D.\ Ball$^3$,
   Alessandro Candido$^{4}$,
   Stefano Carrazza$^5$,
   Juan Cruz-Martinez$^4$,
   Luigi Del Debbio$^3$,\\[0.1cm]
   Stefano Forte$^5$,
   Felix Hekhorn$^{6,7}$,
   Giacomo Magni$^{8,9}$,
   Emanuele R.\ Nocera$^{10}$,\\[0.1cm]
   Tanjona R.\ Rabemananjara$^{8,9}$,
   Juan Rojo$^{8,9}$,
   Roy Stegeman$^3$, and
   Maria Ubiali$^{11}$ 
   
    \vspace{1cm}
    
    {\it \small

    ~$^1$ Deutsches Elektronen-Synchrotron DESY, Notkestr. 85, 22607 Hamburg, Germany\\[0.1cm]
      ~$^2$  Department of Physics and Astronomy, University College London, London, WC1E 6BT, UK\\[0.1cm]
      ~$^3$The Higgs Centre for Theoretical Physics, University of Edinburgh,\\
      JCMB, KB, Mayfield Rd, Edinburgh EH9 3FD, Scotland\\[0.1cm]
      ~$^4$CERN, Theoretical Physics Department, CH-1211 Geneva 23, Switzerland\\[0.1cm]
      ~$^5$Tif Lab, Dipartimento di Fisica, Universit\`a di Milano and\\
      INFN, Sezione di Milano, Via Celoria 16, I-20133 Milano, Italy\\[0.1cm]
      ~$^6$University of Jyvaskyla, Department of Physics, P.O. Box 35, FI-40014 University of Jyvaskyla, Finland\\[0.1cm]
      ~$^7$Helsinki Institute of Physics, P.O. Box 64, FI-00014 University of Helsinki, Finland\\[0.1cm]
      ~$^8$Department of Physics and Astronomy, Vrije Universiteit, NL-1081 HV Amsterdam\\[0.1cm]
      ~$^9$Nikhef Theory Group, Science Park 105, 1098 XG Amsterdam, The Netherlands\\[0.1cm]
      ~$^{10}$Dipartimento di Fisica, Universit\`a degli Studi di Torino and\\
      INFN, Sezione di Torino, Via Pietro Giuria 1, I-10125 Torino, Italy\\[0.1cm]
      ~$^{11}$DAMTP, University of Cambridge, Wilberforce Road, Cambridge, CB3 0WA, United Kingdom\\[0.1cm]
    }
   
    \vspace{2cm}
    
    {\it \small
   
   
  }

{\bf \large Abstract}

\end{center}
We discuss how the two existing approximate N$^3$LO (aN$^3$LO) sets of
parton distributions (PDFs) from the  MSHT20 and NNPDF4.0 series can
be combined for LHC phenomenology, both in the pure QCD case and for
the QCD$\otimes$QED sets that include the photon PDF.
Using the resulting combinations, we present predictions for the total inclusive cross-section for Higgs production in gluon fusion, vector boson fusion, and associated production at the LHC with $\sqrt{s}=13.6$~TeV (Run 3).
For the gluon fusion and vector boson fusion channels, the corrections that arise when using correctly matched aN$^3$LO PDFs with N$^3$LO
cross section calculations, compared to using NNLO PDFs, are significant,
in many cases larger than the PDF uncertainty, and generally larger
than the differences between the two aN$^3$LO PDF sets entering the combination. 
The combined aN$^3$LO PDF sets,  {\tt MSHT20xNNPDF40\_an3lo} and {\tt MSHT20xNNPDF40\_an3lo\_qed}, are made publicly available in the
{\sc\small LHAPDF} format and can be readily used for LHC phenomenology. 

\clearpage

\section{Introduction}
\label{sec:intro}

The two global determinations of parton distribution functions (PDFs) from the MSHT~\cite{Bailey:2020ooq} and NNPDF~\cite{NNPDF:2021njg} collaborations have been recently  extended to approximate next-to-next-to-next-to-leading order (aN$^3$LO), in Refs.~\cite{McGowan:2022nag} and~\cite{NNPDF:2024nan}
respectively. Both the MSHT and NNPDF aN$^3$LO PDF sets have been subsequently enlarged to also
include a photon PDF and mixed QED$\otimes$QCD corrections to
perturbative evolution~\cite{Cridge:2023ryv,Barontini:2024dyb}.  

The inclusion of  N$^3$LO QCD corrections in the
determination of these PDF sets is approximate in two different
respects. First,  N$^3$LO perturbative evolution is only known
approximately. Indeed,
splitting functions are not known exactly,
so an approximation to them
must be constructed based on exact  knowledge of the small- and
large-$x$ behavior at higher perturbative orders (e.g. from resummation) and a finite set of Mellin moments
\cite{Moch:2017uml,Davies:2016jie,Gehrmann:2023cqm,Gehrmann:2023iah,Falcioni:2023tzp,Moch:2018wjh,Moch:2021qrk,Falcioni:2023luc,Falcioni:2023vqq,Moch:2023tdj,Falcioni:2024xyt}. Knowledge of N$^3$LO splitting
functions has progressed rapidly, with more Mellin moments being
gradually published
over the  years.
In fact,
now~\cite{Falcioni:2024xav,Falcioni:2024qpd} the full set
of Mellin moments up to $N=20$ is available for all elements of the
splitting function matrix.
Heavy quark transition matrix elements at $\mathcal{O}(\alpha_s^3)$,
which are needed for massless perturbative evolution in a
variable-flavor number scheme, 
are now also fully known, albeit only recently in complete
form~\cite{Kawamura:2012cr,Bierenbaum:2009mv,Ablinger:2014vwa,Ablinger:2014nga,Blumlein:2021enk,Ablinger:2014uka,Ablinger:2014tla,ablinger:agq,Ablinger:2022wbb,Ablinger:2023ahe,Ablinger:2024xtt}. Hence
N$^3$LO perturbative evolution is known exactly for all practical
purposes, as we shall discuss in somewhat more detail below.

Second, however,
N$^3$LO partonic cross-sections are only known for a subset of the processes used for PDF determination, namely  massless
deep-inelastic scattering (DIS)\cite{Vermaseren:2005qc,Blumlein:2022gpp}  and the  total Drell-Yan cross-section
and on-shell rapidity distribution\cite{Caola:2022ayt,Duhr:2020sdp,Duhr:2020seh,duhr:DY2021,Gehrmann:DYN3LO}. In particular, whereas
the computation of the high $Q^2$ limit of
heavy quark structure functions was recently
completed, permitting the determination of
heavy quark transition matrix elements, the computation of corrections
to DIS coefficient functions
proportional to powers of $\frac{m_h^2}{Q^2}$ of the heavy quark
mass is beyond the current state of the art, as is the computation
of fiducial N$^3$LO corrections to hadronic processes, which are
needed for a meaningful comparison to hadron collider data.

As a consequence, while approximate N$^3$LO perturbative evolution  is
close to exact, knowledge of N$^3$LO hard cross sections is  limited.
Moreover, massless DIS data do not fully determine the PDFs, and
consequently a N$^3$LO PDF determination must include a number of
processes for which only NNLO matrix elements are available.
It is in this respect that perturbative accuracy in aN$^3$LO PDF determination is not fully N$^3$LO.

The two available aN$^3$LO PDF determinations include estimates of the uncertainty
due to incomplete knowledge of N$^3$LO terms, namely, both the incomplete
knowledge of splitting functions, and the missing N$^3$LO hard
cross-sections,
as well as that related
to missing perturbative corrections at N$^4$LO and beyond.
In Ref.~\cite{McGowan:2022nag} (MSHT) the uncertainty estimation is done by
means of nuisance parameters, while in Ref.~\cite{NNPDF:2024nan} (NNPDF) it is done
through a theory covariance matrix. 
As is well known, (see
e.g.\ Refs.~\cite{Ball:2012wy,Ball:2021icz}) the nuisance parameter and theory covariance matrix approaches are completely equivalent for Gaussian uncertainties, provided only the eigenvectors and eigenvalues of the covariance matrix coincide with the nuisance parameters and their uncertainties. 

The approaches of Refs.~\cite{McGowan:2022nag,NNPDF:2024nan} consequently
only differ in the way the covariance matrix, or the corresponding 
nuisance parameters, are estimated. In Ref.~\cite{McGowan:2022nag} this
is done based on a judicious estimate of the form of missing terms,
which
also incorporates effects of varying the parametrization of the
splitting functions. 
 In Ref.~\cite{NNPDF:2024nan}, instead,  the contributions to the
 covariance matrix due to incomplete knowledge of splitting functions
 are estimated differently from
 all those due to missing higher order terms (missing N$^4$LO
 splitting functions, missing N$^4$LO 
 massless DIS coefficient function, and missing N$^3$LO hard cross
 sections for all other processes). Namely,
 variation of the
 parametrization of splitting functions (as in
 Ref.~\cite{McGowan:2022nag}) is used for the estimate of
 their incomplete knowledge, while renormalization and factorization scale variation is used for missing
higher order perturbative corrections.

Both Refs.~\cite{McGowan:2022nag,NNPDF:2024nan} included all the
information that was available at the time of their
publication, and consequently, because Ref.~\cite{McGowan:2022nag} was
published earlier, it included less information on N$^3$LO evolution,
and specifically it did not yet include the more recent information on Mellin moments from
Refs.~\cite{Davies:2022ofz,Falcioni:2023luc,Falcioni:2023vqq,Moch:2023tdj,
  Falcioni:2023tzp}, and the exact knowledge of the heavy
quark transition matrix elements, both of which are instead included
in Ref.~\cite{NNPDF:2024nan} (with the only exception of the heavy quark terms of Ref.~\cite{Ablinger:2024xtt}) as they had become available
meanwhile. Neither determination includes the higher moments of 
$P_{gq}$~\cite{Falcioni:2024xyt} and
$P_{gg}$~\cite{Falcioni:2024xav,Falcioni:2024qpd}. Some
uncertainties in  Ref.~\cite{McGowan:2022nag} are accordingly larger, but
with the information now available it is possible
to check that uncertainties in
Refs.~\cite{McGowan:2022nag,NNPDF:2024nan} were estimated accurately.

Specifically, the uncertainty on splitting functions with Mellin
moments up to $N=20$ for all matrix
elements but $P_{gg}$ was assessed in both
Refs.~\cite{Moch:2023tdj,NNPDF:2024nan} with very similar results, and
shown in Ref.~\cite{NNPDF:2024nan} to be smaller than the uncertainty
due to the missing  N$^4$LO
corrections estimated by scale variation. This assessment was extended
to $P_{gg}$ recently in Ref.~\cite{Falcioni:2024qpd}. The only other
uncertainty on perturbative evolution present in
Ref.~\cite{NNPDF:2024nan} are the contributions, computed in Ref.~\cite{Ablinger:2024xtt}, to transition
matrix elements, which however were found in Ref.~\cite{Ablinger:2024xtt} to agree perfectly within the uncertainty band  with the previous approximation~\cite{Kawamura:2012cr} adopted in Ref.~\cite{NNPDF:2024nan}. Based on this, it can be safely concluded
that, as mentioned, N$^3$LO perturbative evolution is fully known for all
practical purposes. Furthermore,  the splitting 
functions  of  Refs.~\cite{McGowan:2022nag,NNPDF:2024nan} agree within
uncertainties, with the exception of the $P_{gq}$ splitting function,
that differs at approximately the two-sigma level, see also the benchmarking comparison of~\cite{Andersen:2024czj,Cooper-Sarkar:2024crx}. Note, however, that as far as PDF evolution is concerned $P_{gg}$ is much more important than $P_{gq}$, and the effects at small-$x$ are to a large extent anti-correlated. Updated knowledge of the splitting functions effectively removes this one source of moderate discrepancy.  

A preliminary analysis of the effect of including these additional
Mellin moments, and hence improved splitting functions, which have been determined
since the original MSHT publication was presented
in~\cite{Thorne:2024npj}, and observed a small increase in the gluon
near $x=0.01$, though still within the original MSHT uncertainties. While this 
is significantly washed out in parton luminosities, it does further improve the agreement between the two PDF sets. Furthermore, updated PDF sets obtained using a parametrization of the N$^3$LO splitting function constructed in~\cite{Falcioni:2024qpd} that includes the full  information on Mellin moments (and most recently by MSHT, also the complete heavy quark transition matrix elements already used by NNPDF) now available have been presented in preliminary form  both by  MSHT~\cite{mshttalk,mshttalk2} and NNPDF~\cite{nnpdftalk}. These preliminary results allow for a comparison in which all the currently available  information on perturbative evolution is consistenly used by the two groups. Based on these results, it can be concluded that,  whereas the use of a common parametrization of splitting functions and transition matrix elements reduces somewhat the differences between the MSHT and NNPDF aN$^3$LO PDF sets, results change by a small amount. We shall come back to this point in Sect.~\ref{sec:higgs} below.

Estimates of the impact of the use of  aN$^3$LO PDFs in the
computation of the Higgs total production cross-section in both
Refs.~\cite{McGowan:2022nag,NNPDF:2024nan} suggest that  using NNLO PDFs, instead of aN$^3$LO PDFs,  with the N$^3$LO matrix
element for Higgs production  in gluon fusion
and vector boson fusion 
leads to an error that is comparable to, or even larger than the
PDF uncertainty on the N$^3$LO result. Also, this error
is found~\cite{NNPDF:2024nan} to be rather larger than the estimate of
its size
based on the corresponding shift at one less perturbative order~\cite{Baglio:2022wzu,Anastasiou:2016cez}. 
This suggests that using NNLO PDFs in the  N$^3$LO computation spoils the accuracy of the  N$^3$LO result, in a way that, moreover, may not be reflected in a corresponding increase in uncertainty, if the latter is estimated in this way.

Given the availability of two different  aN$^3$LO PDF sets, accurate  N$^3$LO phenomenology requires therefore a comparison of the impact of the use of either of these PDF sets, and a prescription for their combination. 
It is the purpose of this brief note to provide a first assessment and suggest a possible prescription, having specifically in view  Higgs
production in gluon fusion, but presenting results also for other Higgs production modes.
We will first compare  the aN$^3$LO PDFs
of Refs.~\cite{McGowan:2022nag,NNPDF:2024nan}, both without and with QED corrections, to each other, then assess for each of them  the impact of aN$^3$LO corrections on parton luminosities, and finally construct a combination of these PDFs based on the PDF4LHC combination methodology~\cite{Butterworth:2015oua}. 
We will finally discuss results for the cross-section for Higgs production in gluon fusion and other channels based on individual and combined aN$^3$LO PDF sets.

\section{Comparison of aN$^3$LO PDFs}
\label{sec:comp}

We will compare NNPDF4.0 and MSHT PDF sets both with and without QED effects. The QED sets include a photon PDF that mixes with the quark and gluon PDFs upon combined QED$\times$QCD evolution, and mostly differ from the pure QCD set due to slight depletion of the gluon PDF due to the momentum fraction carried by the photon. They are consequently important when assessing the behavior of the gluon PDF.

When comparing MSHT and
NNPDF PDF sets, care must be taken to compare sets that include the
same uncertainties.
Specifically, all NNPDF4.0 PDF sets are now
delivered in two different versions, one in which the PDF uncertainty
includes  the uncertainty due to missing higher-order corrections on
the theory predictions used for PDF determination (MHOU, henceforth),
and the other in 
which it does not. Hence uncertainties due to missing  N$^4$LO
corrections are only included by NNPDF in the aN$^3$LO MHOU sets, though
both NNPDF and MSHT aN$^3$LO sets include uncertainties due to missing
N$^3$LO corrections to hadronic processes and massive DIS. Note
that the NLO and NNLO sets with MHOU were not
included at the time of first release of NNPDF4.0~\cite{NNPDF:2021njg}, and only
released at  a later stage~\cite{NNPDF:2024dpb}. 
 On the other hand, in the MSHT approach all
uncertainties, including all MHOUs, are included at N$^3$LO, while MHOUs
are not included at NNLO (as they were not in the original NNLO
NNPDF4.0 PDFs of Ref.~\cite{NNPDF:2021njg}).
The pertinent comparison of the MSHT20 sets
is therefore at N$^3$LO with the NNPDF4.0 sets with MHOUs
included, while at  NNLO with the NNPDF4.0 sets without MHOUs. Note that in Ref.~\cite{NNPDF:2024nan} the MSHT20 aN$^3$LO PDFs were instead compared to the NNPDF4.0 set without MHOUs.  Henceforth we will here tacitly assume that for
NNPDF4.0 NNLO refers to the non-MHOU sets while aN$^3$LO refers to the
MHOU sets. 

When comparing the aN$^3$LO PDF sets, it should of course be borne in
mind that the MSHT20 and NNPDF4.0 sets already differ at NNLO due to differences in
dataset and methodology. On top of the difference in basic underlying
methodology, namely generalized polynomial parametrization and Hessian
uncertainties for MSHT, neural network parametrization and Monte Carlo
uncertainties for NNPDF, several details of the theoretical approach
also differ. Specifically, there  are several differences in the
treatment of heavy quarks: the values of heavy quark masses are
somewhat different; the charm PDF is fitted to the
data in NNPDF4.0, while in MSHT20 it is determined by perturbative
matching conditions; and for DIS the massive and massless heavy quark schemes
are matched using the Thorne-Roberts scheme by MSHT and the FONLL
scheme by NNPDF.
A detailed comparison of the NNLO MSHT20 and
NNPDF4.0 PDF sets was performed in  Ref.~\cite{NNPDF:2021njg}, see in
particular Fig.~21 and Fig.~60 of that reference.  
Several of the differences seen in the comparison of Ref.~\cite{NNPDF:2021njg} are also seen when comparing
MSHT20 to the previous NNPDF3.1 PDF set, and indeed many of the methodological differences 
(in particular those related to the
treatment of heavy quarks) are the same. This is relevant because MSHT20 and NNPDF3.1  were
included in a detailed benchmarking 
presented in Ref.~\cite{PDF4LHCWorkingGroup:2022cjn,Cridge:2021qjj},
along with the CT18 PDF set. 
In this work,  dedicated PDF sets from the three groups were produced by
adopting common assumptions and settings (specifically for the treatment of heavy quarks), instead of the inconsistent settings used in the original fits, as well as a relatively similar
global dataset. These PDF sets were called ``reduced" in Ref.~\cite{PDF4LHCWorkingGroup:2022cjn} (though not based on a reduced dataset).
These dedicated sets turned out to be in quite good agreement with each other within uncertainties, thereby showing that the differences between the original sets were partly due to these inconsistent settings.  

These more consistent  sets were then used to produce the PDF4LHC21 combination~\cite{PDF4LHCWorkingGroup:2022cjn}. The combination that we
will discuss  here is instead based on  the
publicly available aN$^3$LO PDF sets, thereby leading to a more
conservative uncertainty estimate that encompasses the differences
between the PDF sets entering the combination, as we shall explain
below. To understand the features of this combination
we therefore start with a  comparison of the published MSHT20 and
NNPDF4.0 aN$^3$LO sets, along with the NNLO sets, to see how
much of the differences seen at    aN$^3$LO
reflect those already present at NNLO, whose origin was largely understood in  Refs.~\cite{PDF4LHCWorkingGroup:2022cjn,Cridge:2021qjj}.


\begin{figure}[!p]
  \centering
  \includegraphics[width=0.45\textwidth]{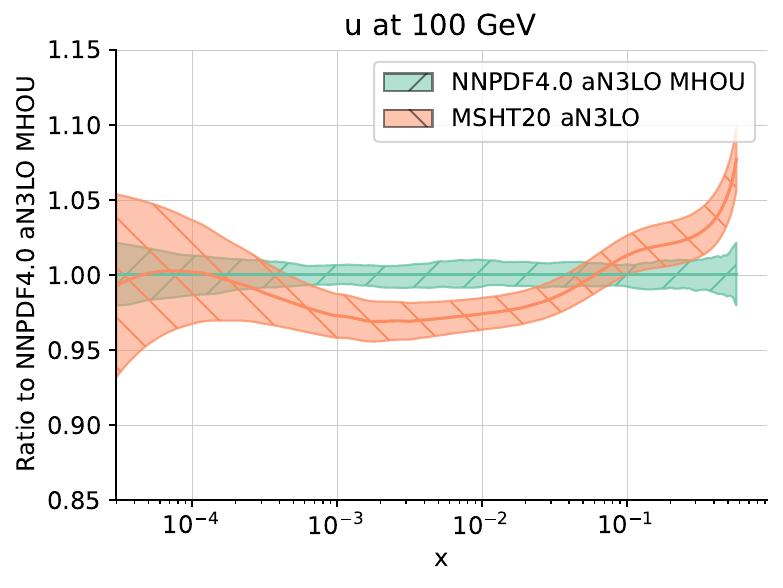}
  \includegraphics[width=0.45\textwidth]{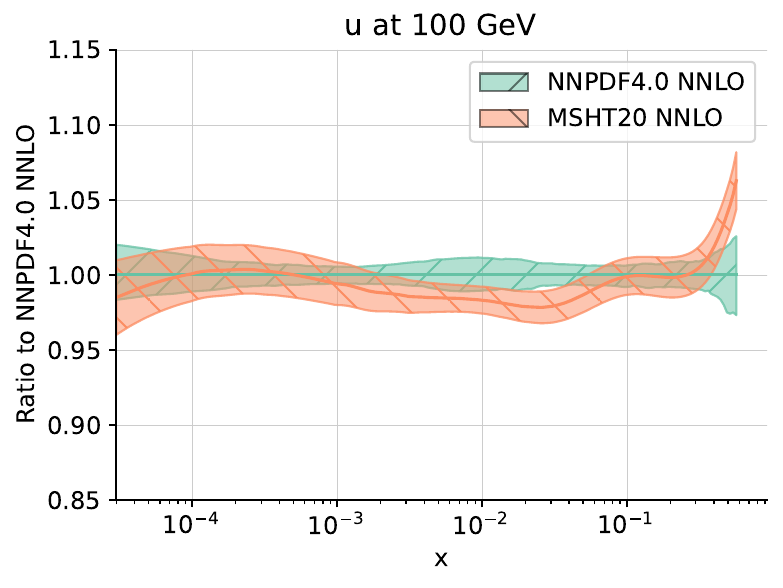}\\
  \includegraphics[width=0.45\textwidth]{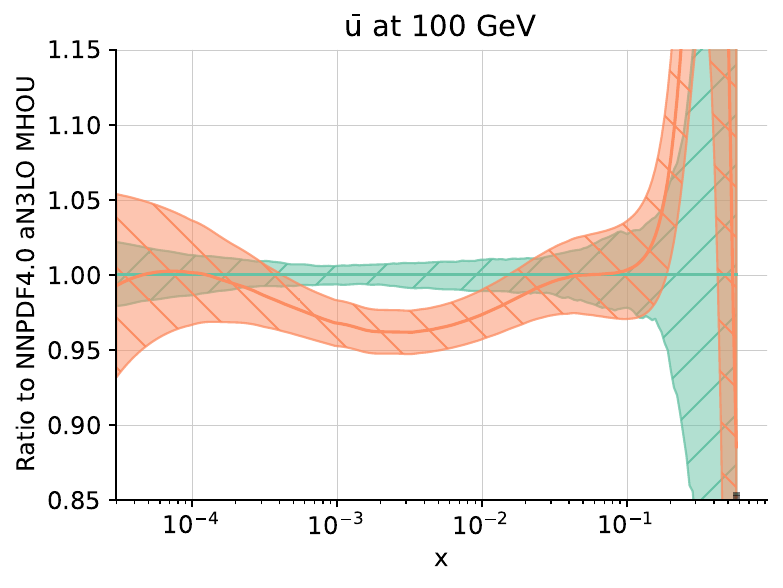}
  \includegraphics[width=0.45\textwidth]{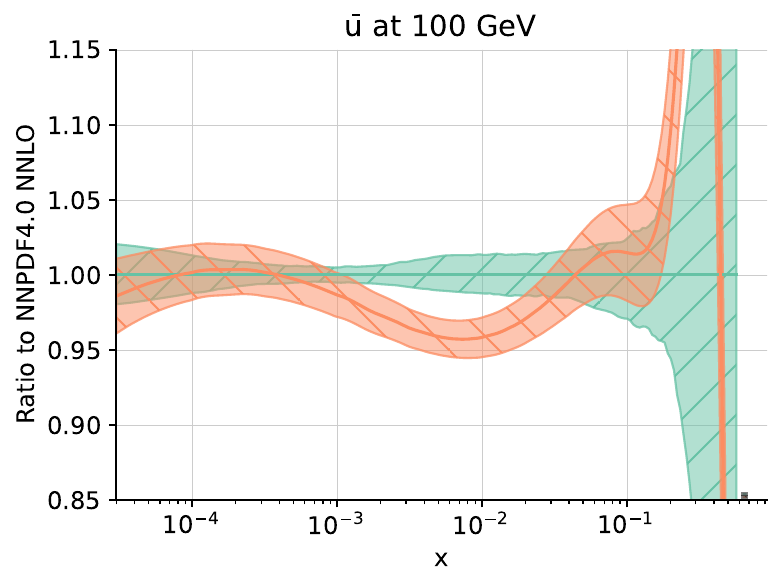}\\
  \includegraphics[width=0.45\textwidth]{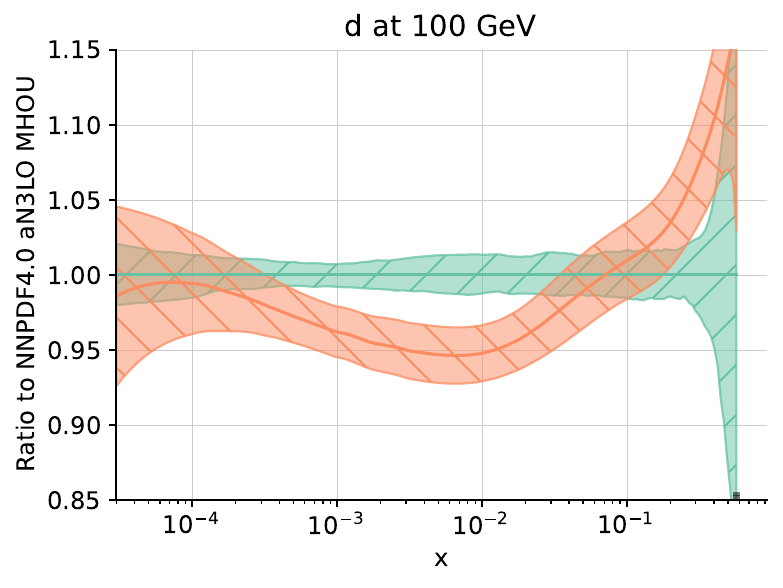}
  \includegraphics[width=0.45\textwidth]{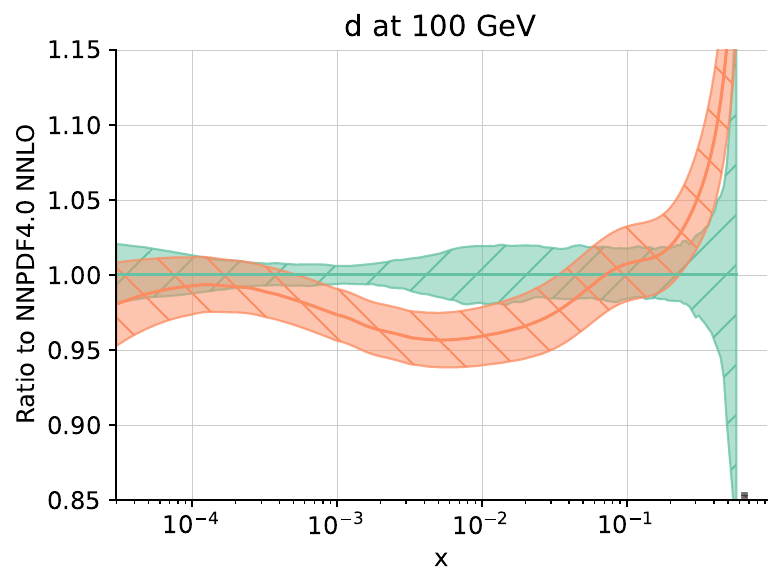}\\
  \includegraphics[width=0.45\textwidth]{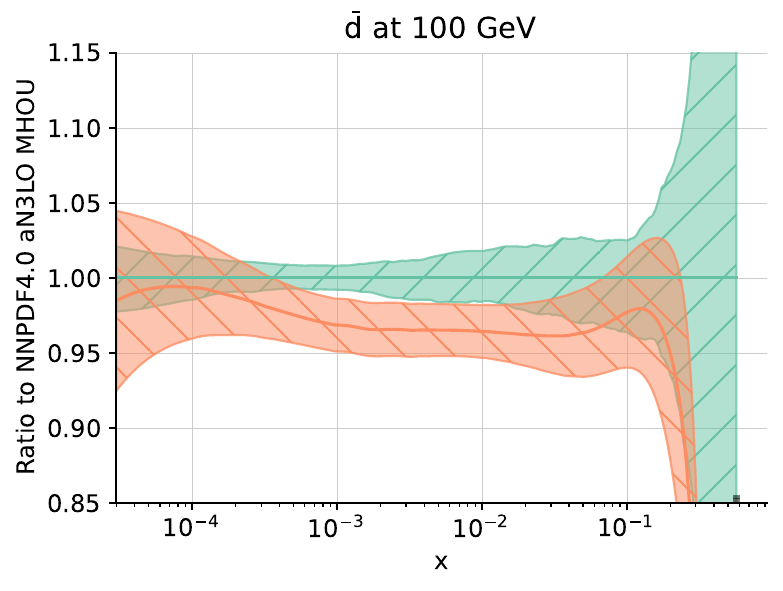}
  \includegraphics[width=0.45\textwidth]{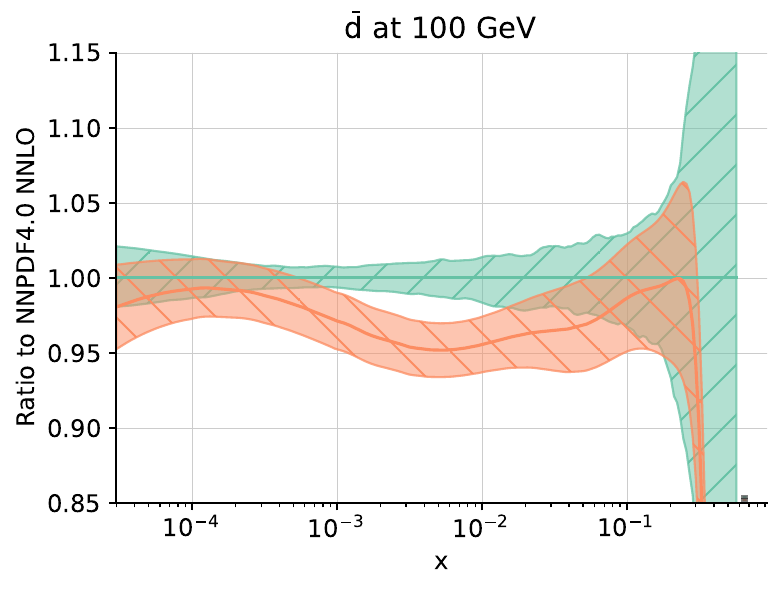}\\
  \caption{The NNPDF4.0~\cite{NNPDF:2024nan} and MSHT20~\cite{McGowan:2022nag} aN$^3$LO (left) and NNLO (right) PDFs, for the $u$ and $d$ sector at $Q=100$~GeV, shown as a ratio to NNPDF4.0. All uncertainties shown
  are one sigma.}
  \label{fig:PDFs_N3LO_vs_NNLO_1}
\end{figure}

\begin{figure}[!p]
  \centering
  \includegraphics[width=0.45\textwidth]{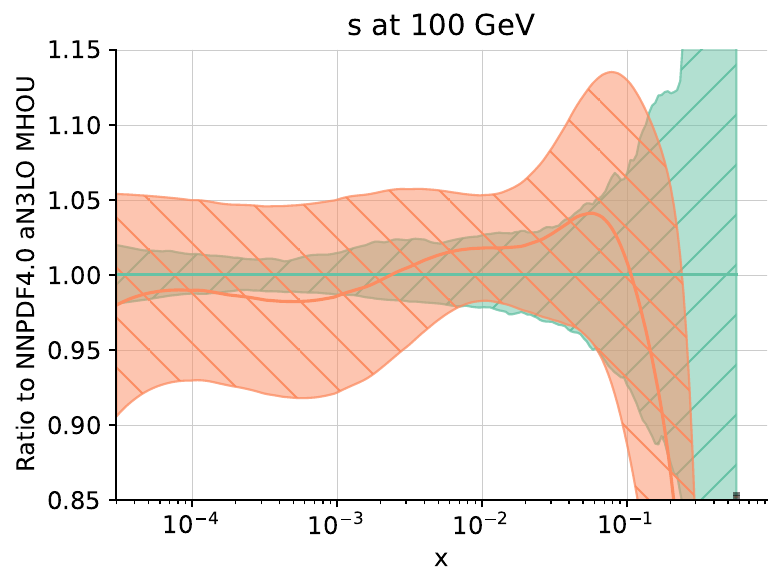}
  \includegraphics[width=0.45\textwidth]{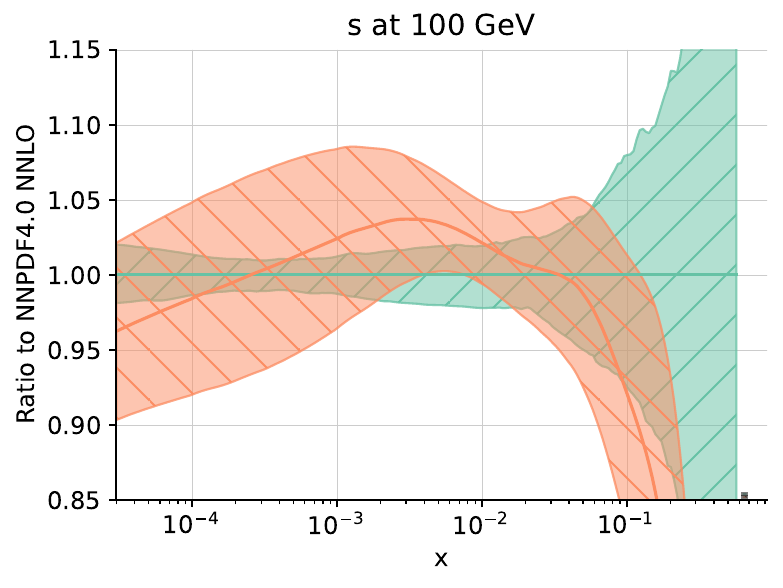}\\
  \includegraphics[width=0.45\textwidth]{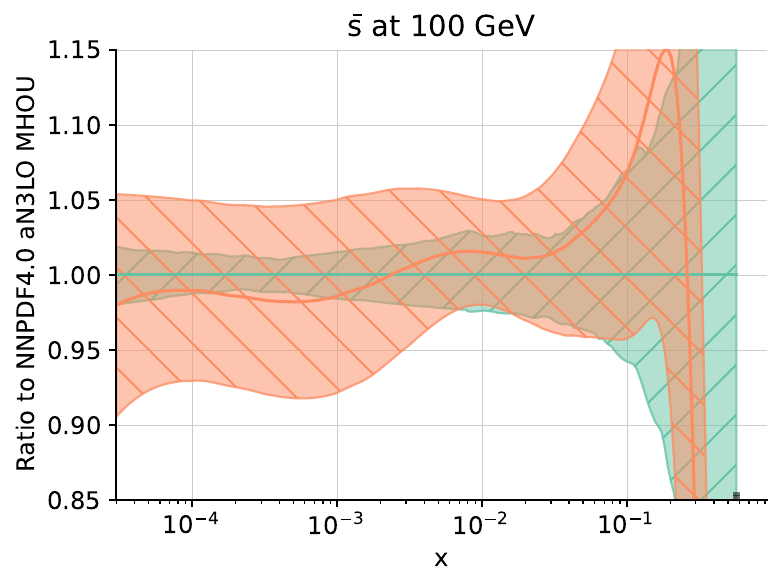}
  \includegraphics[width=0.45\textwidth]{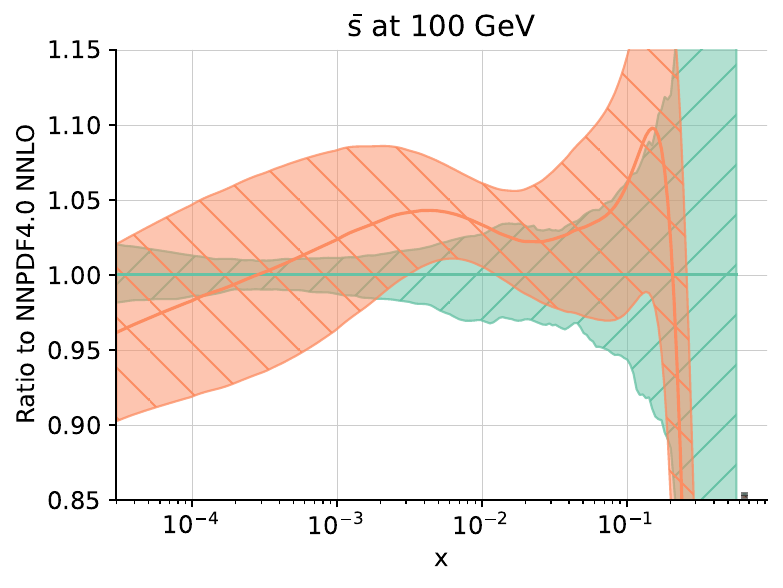}\\
  \includegraphics[width=0.45\textwidth]{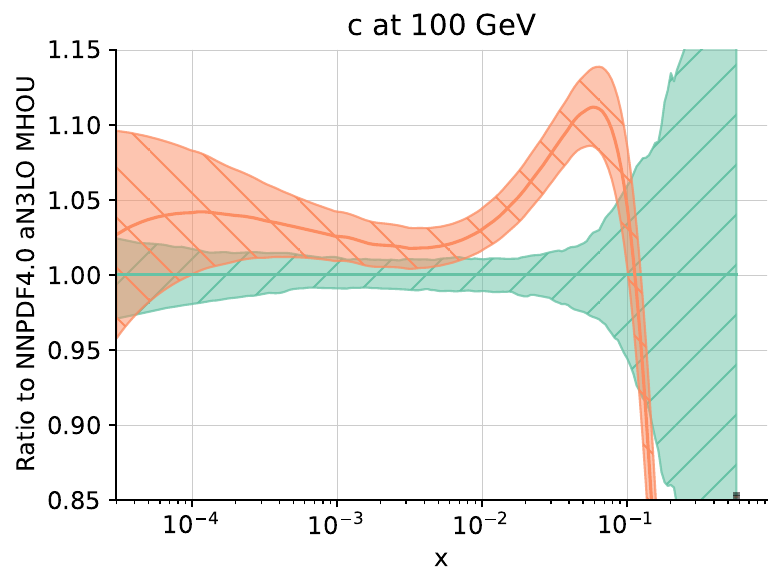}
  \includegraphics[width=0.45\textwidth]{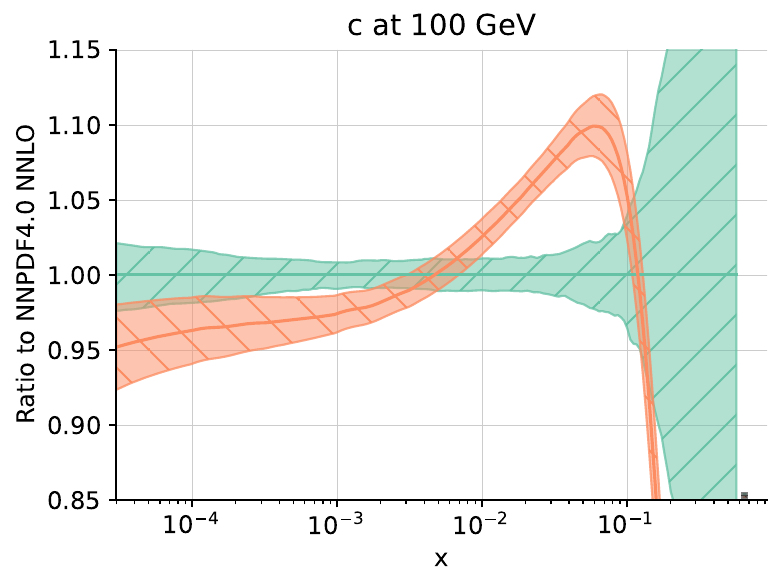}\\
  \includegraphics[width=0.45\textwidth]{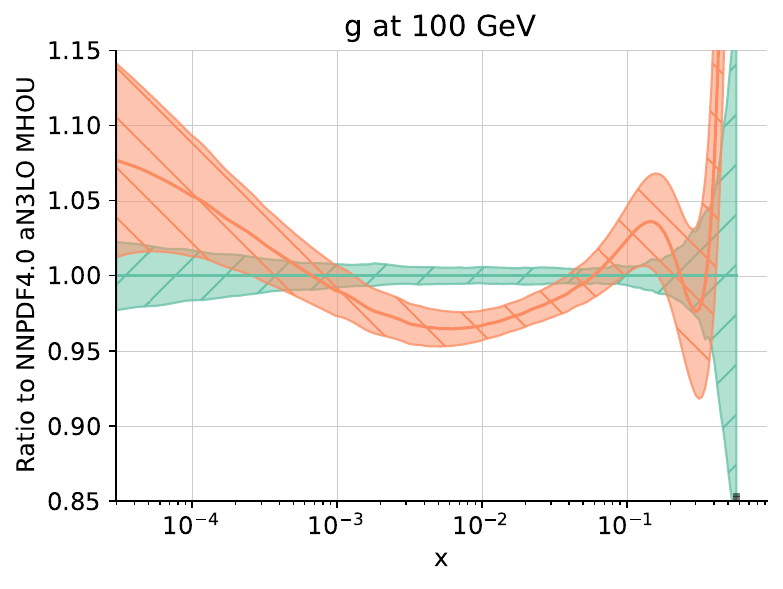}
  \includegraphics[width=0.45\textwidth]{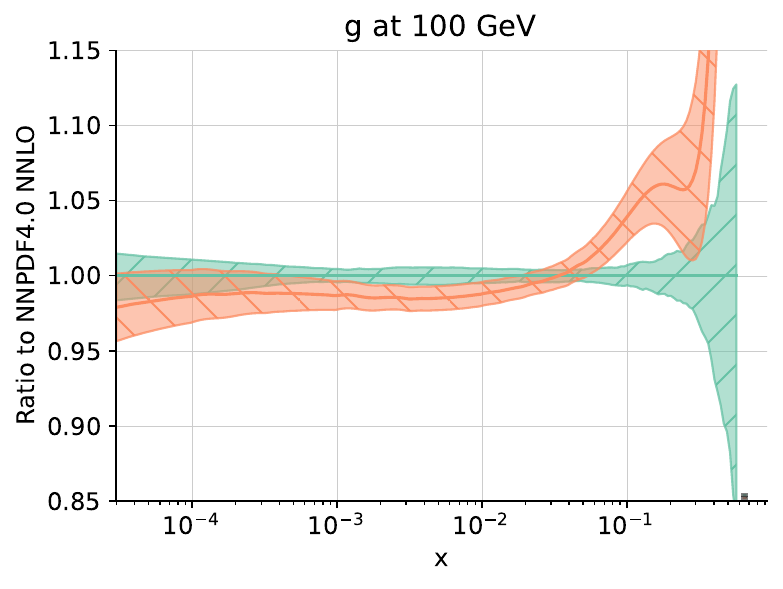}\\
  \caption{As in Fig.~\ref{fig:PDFs_N3LO_vs_NNLO_1} but for the $s,\overline{s},c,g$ PDFs.}
  \label{fig:PDFs_N3LO_vs_NNLO_2}
\end{figure}



\begin{figure}[!p]
  \centering
  \includegraphics[width=0.45\textwidth]{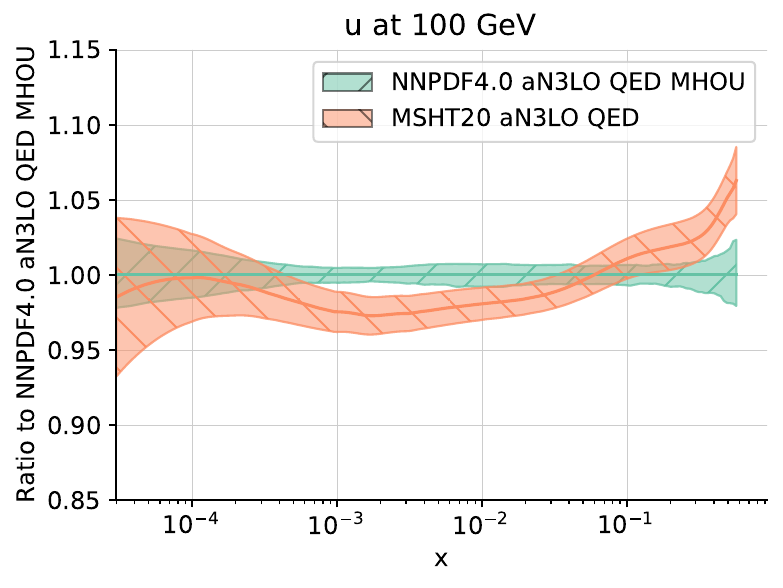}
  \includegraphics[width=0.45\textwidth]{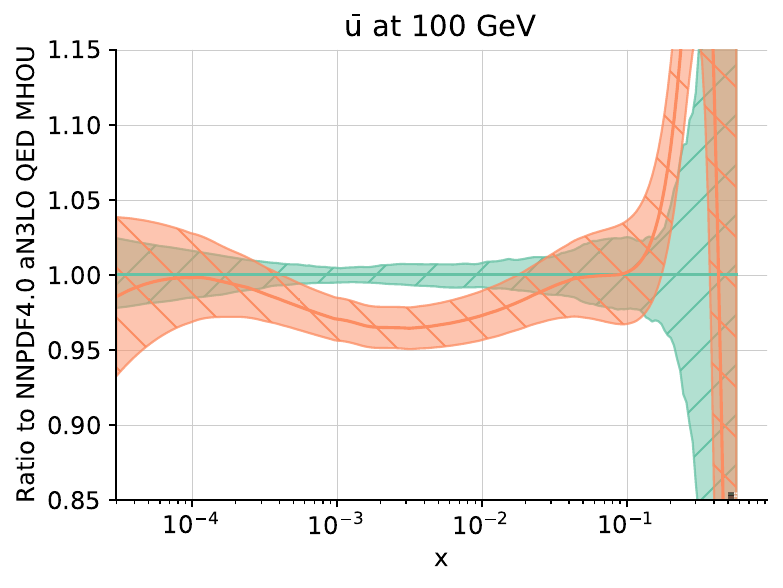}\\
  \includegraphics[width=0.45\textwidth]{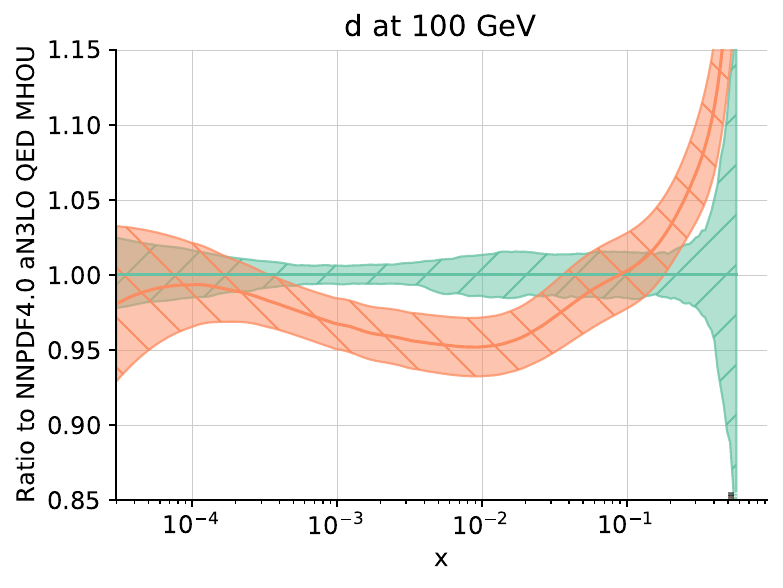}
  \includegraphics[width=0.45\textwidth]{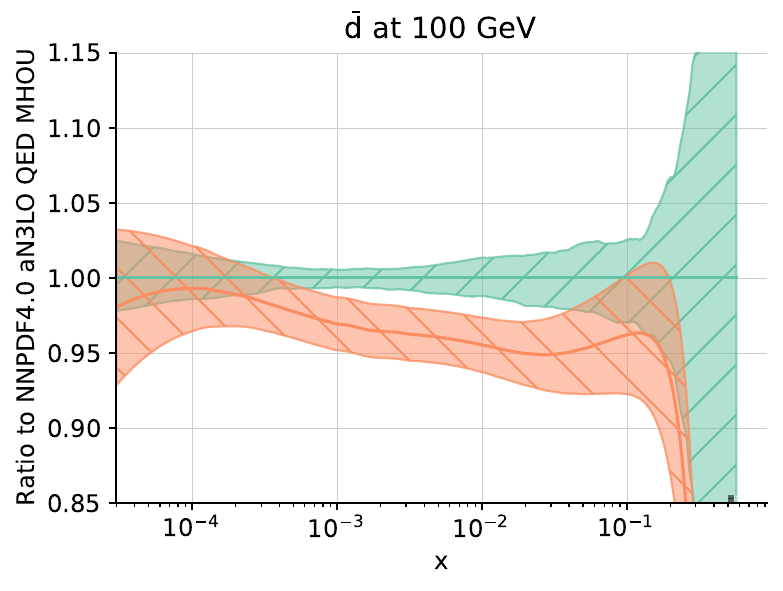}\\
  \includegraphics[width=0.45\textwidth]{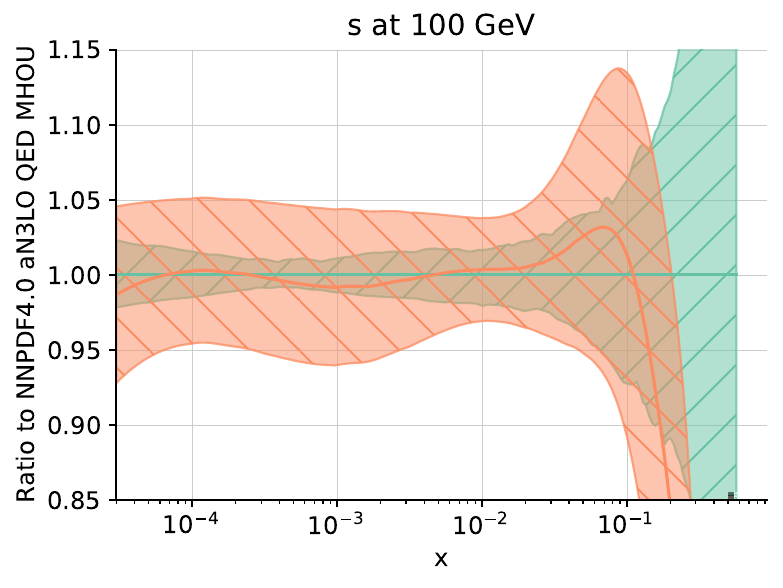}
  \includegraphics[width=0.45\textwidth]{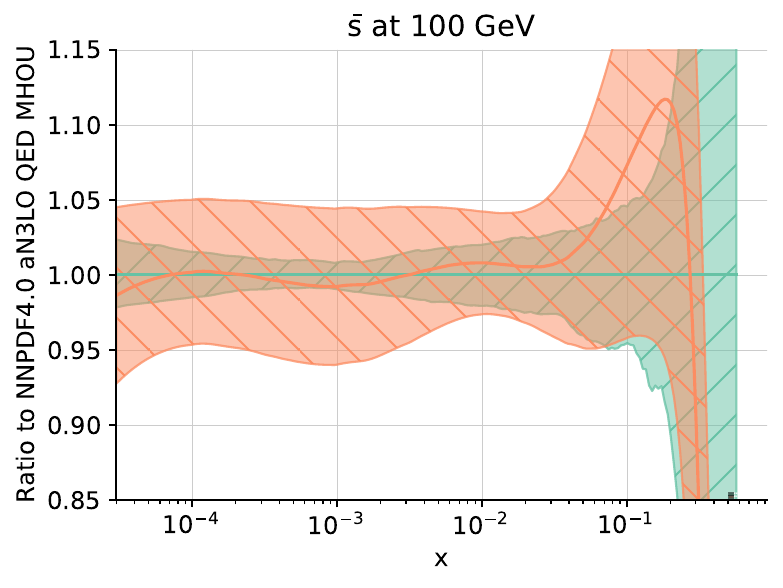}\\
  \includegraphics[width=0.45\textwidth]{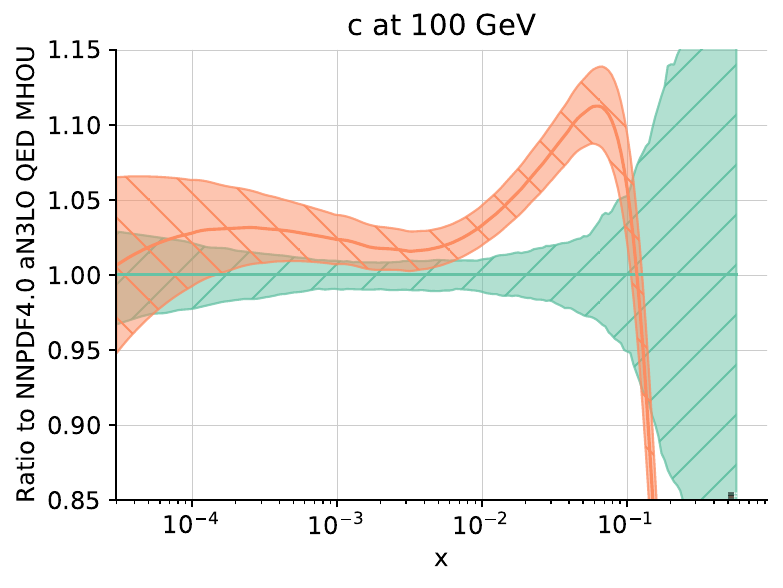}
  \includegraphics[width=0.45\textwidth]{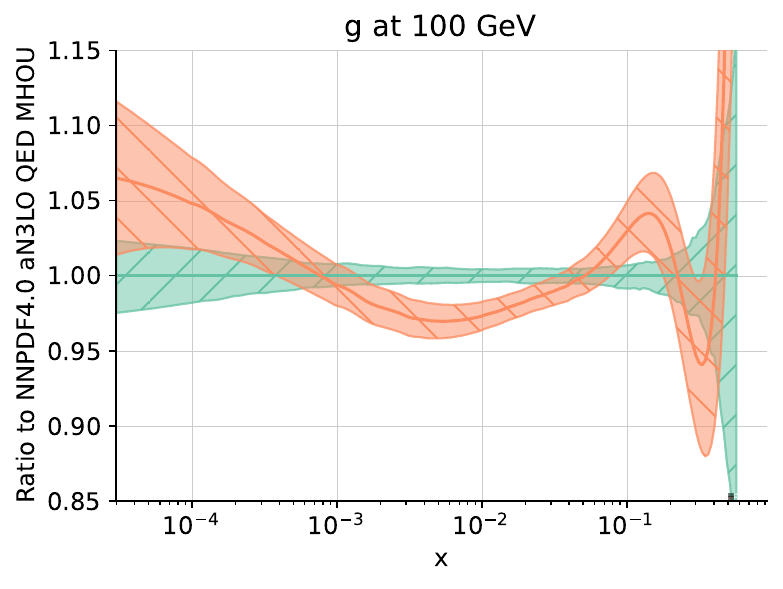}\\
\caption{The NNPDF4.0 ~\cite{Barontini:2024dyb} and MSHT20~\cite{Cridge:2023ryv} aN$^3$LO+QED PDFs at $Q=100$~GeV, shown as a ratio to NNPDF4.0. All uncertainties shown are one sigma.}
  \label{fig:PDFs_MSHT20_QED}
\end{figure}


The comparison of the aN$^3$LO sets is presented in
Figs.~\ref{fig:PDFs_N3LO_vs_NNLO_1}-\ref{fig:PDFs_N3LO_vs_NNLO_2}
(left), together with the comparison of the corresponding NNLO sets (right).  The  aN$^3$LO sets including QED corrections  are compared in Fig.~\ref{fig:PDFs_MSHT20_QED}.
All  PDFs are shown at $Q=100$~GeV,
normalized to the NNPDF4.0 central 
value. All error bands are  one sigma uncertainties.  The dominant differences between the aN$^3$LO PDF sets are 
essentially the same as at NNLO, with the largest difference
observed for the charm PDF, which, as mentioned,  is independently parametrized in
NNPDF4.0, but not in MSHT20, where it is determined by perturbative
matching conditions. Indeed, it is clear from
Figs.~\ref{fig:PDFs_N3LO_vs_NNLO_1}-\ref{fig:PDFs_N3LO_vs_NNLO_2}
that most 
differences between MSHT and NNPDF change very little   when moving from NNLO to aN$^3$LO.
An exception is  the gluon PDF, which differs more between NNPDF and MSHT at
aN$^3$LO. Specifically, 
while reasonably compatible for 
$x\lesssim 0.07$ at NNLO, the MSHT and NNPDF gluons disagree  at aN$^3$LO, with the
MSHT20 result suppressed in comparison to NNPDF4.0 by 3-4\% in the region
$10^{-3}\lesssim x \lesssim 10^{-1}$, with a PDF uncertainty of
1-2\%.

This suppression can, to some extent, be
traced to the difference in the behavior of  $P_{gq}$ mentioned in the
introduction.  The  $P_{gq}$ splitting function contains more unknown small-$x$ divergent 
terms than the other splitting functions, and hence has more potential uncertainty at small $x$, especially if a color-charge relationship that would relate these unknown terms to known contributions to $P_{gg}$ is not assumed to be approximately true for subleading terms (and indeed it is assumed by NNPDF but not by MSHT). Hence, with the smaller moment information and weaker constraints used by MSHT~\cite{McGowan:2022nag}, deviations are possible but have been largely eliminated with the improved information now available. 

In practice, the differences between the two aN$^3$LO PDF sets
are at most at the two sigma level.
Furthermore, a recent MSHT preliminary study shows  that updating the
splitting function with the additional moments now
known~\cite{Falcioni:2023luc,Falcioni:2023vqq,Moch:2023tdj} results in
a $\sim 1.5\%$ increase in the gluon at $x \sim 0.01$~\cite{Thorne:2024npj}, though it does still remain within the uncertainty band
of the original MSHT determination, thereby further reducing
differences. Hence, updates in $P_{gq}$ do improve the comparison, but do not remove the difference in the gluon. Recall however that,  as mentioned in Sect.~\ref{sec:intro}, the very recent
additional moments now known for $P_{gq}$~\cite{Falcioni:2024xyt} and $P_{gg}$~\cite{Falcioni:2024xav,Falcioni:2024qpd}  are not included in either the MSHT or NNPDF sets discussed here. Finally, the inclusion of the photon PDF also has a moderate impact on the other
PDFs, mostly limited to a suppression of the gluon PDF due to
the 1\% shift in momentum fraction from the gluon to the photon, as
well as  some reduction in the $u,\overline{u}$ PDFs at high $x$,
due to $q\to q\gamma$ splitting.

\begin{figure}[!t]
  \centering
  \includegraphics[width=0.45\textwidth]{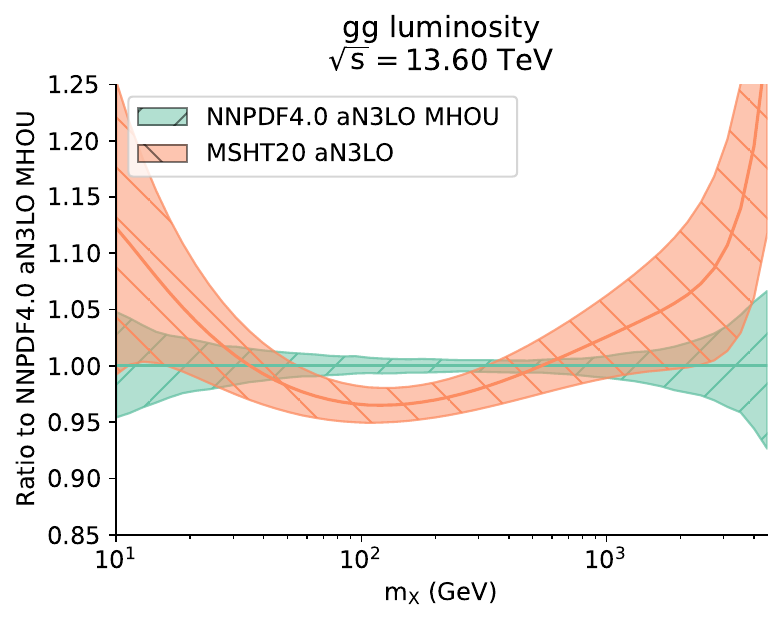}
  \includegraphics[width=0.45\textwidth]{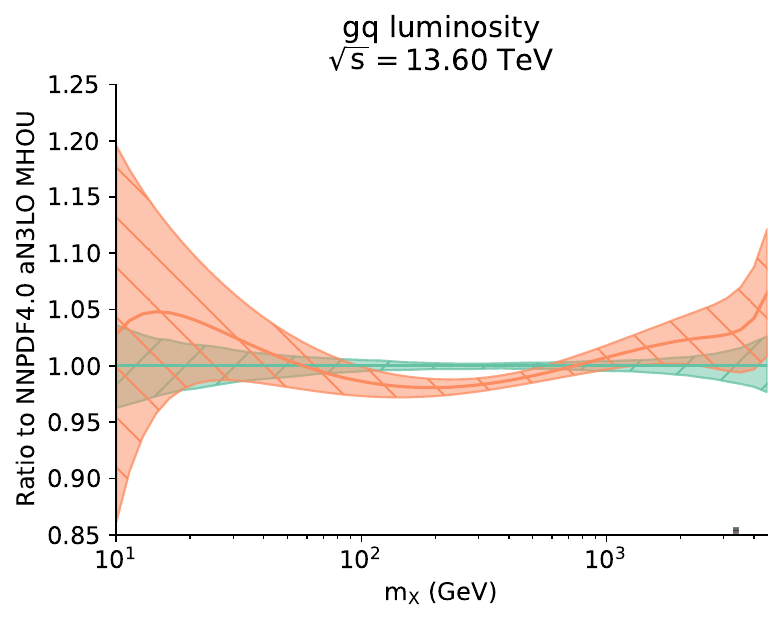}\\
  \includegraphics[width=0.45\textwidth]{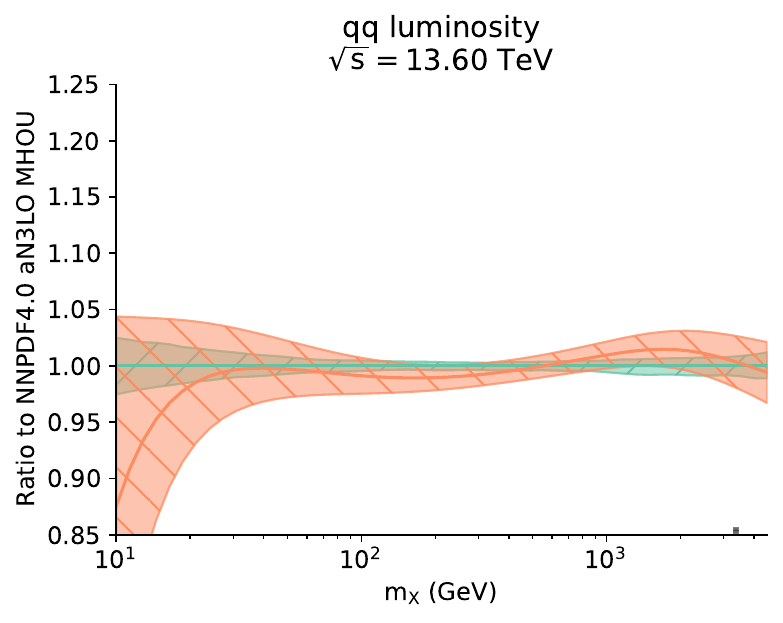}
  \includegraphics[width=0.45\textwidth]{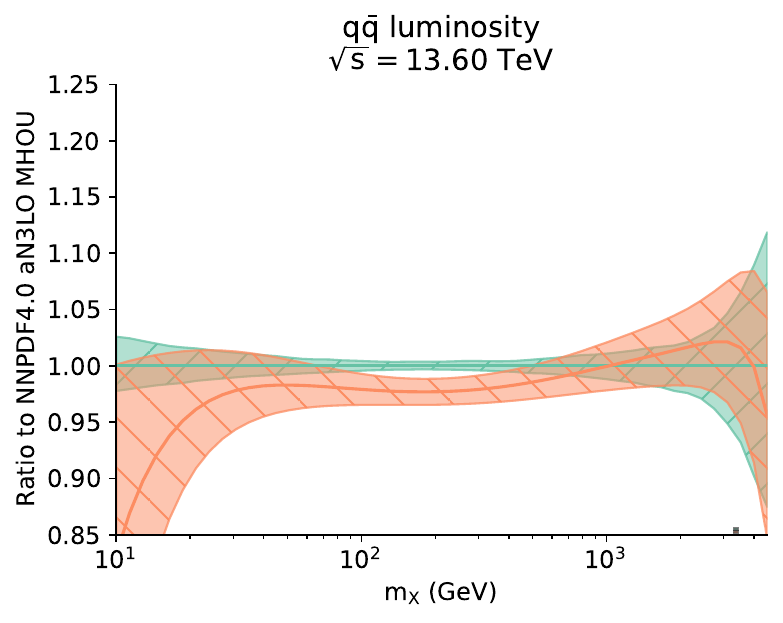}\\
  \caption{The same comparison as in Figs.~\ref{fig:PDFs_N3LO_vs_NNLO_1}-\ref{fig:PDFs_N3LO_vs_NNLO_2} (right) but for the corresponding parton luminosities.}
  \label{fig:lumis_MSHT20}
\end{figure}

\begin{figure}[!t]
  \centering
  \includegraphics[width=0.45\textwidth]{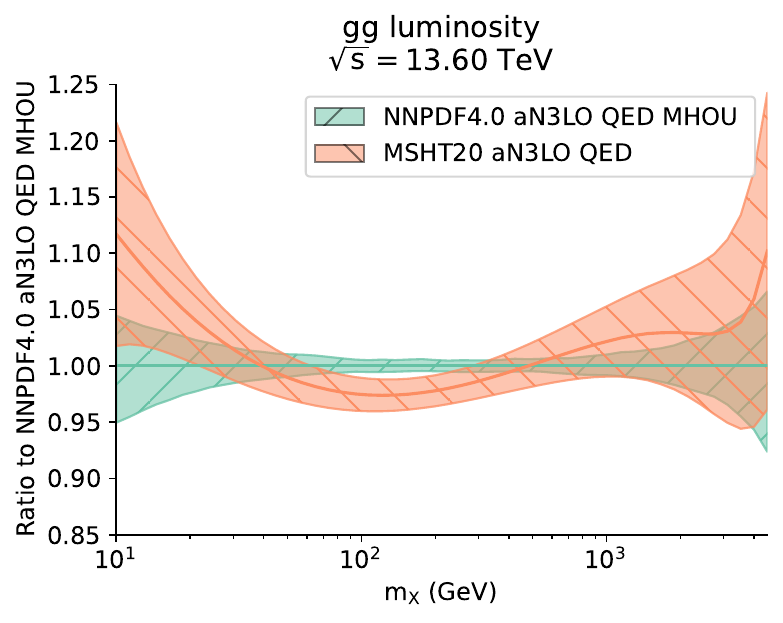}
  \includegraphics[width=0.45\textwidth]{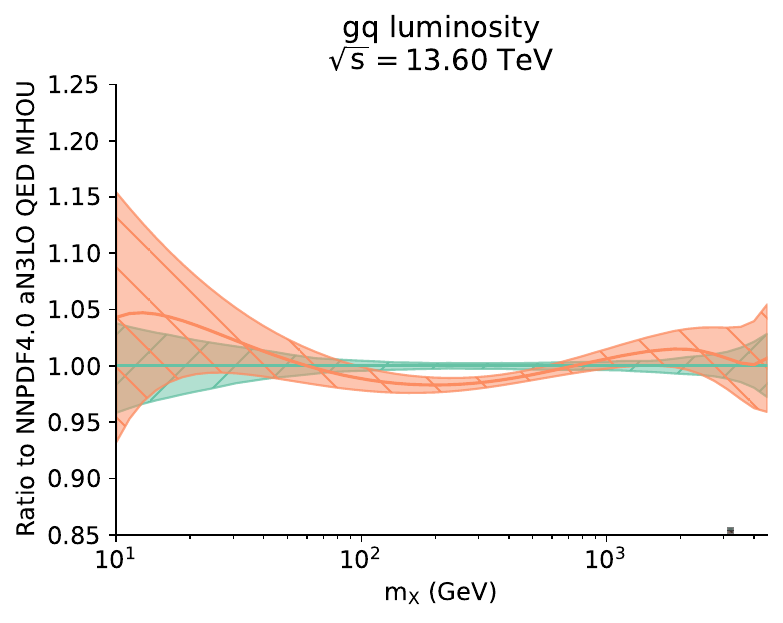}\\
  \includegraphics[width=0.45\textwidth]{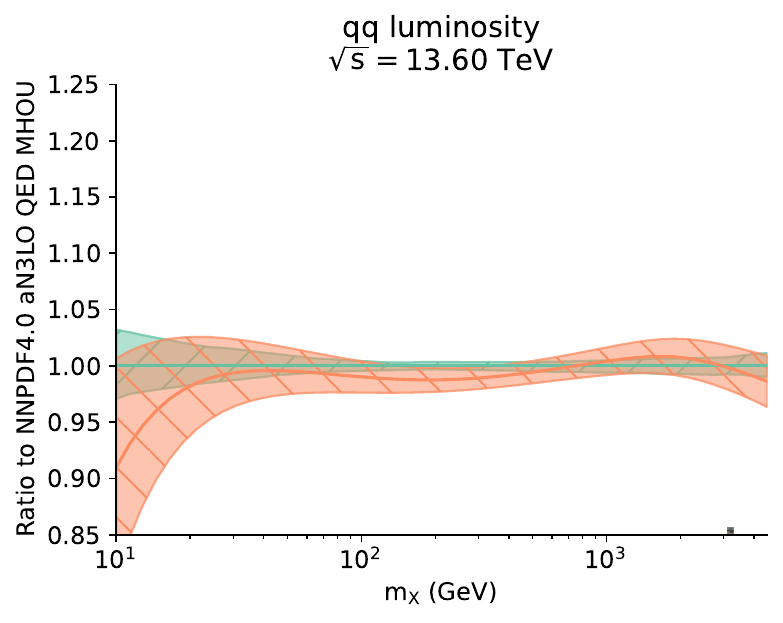}
  \includegraphics[width=0.45\textwidth]{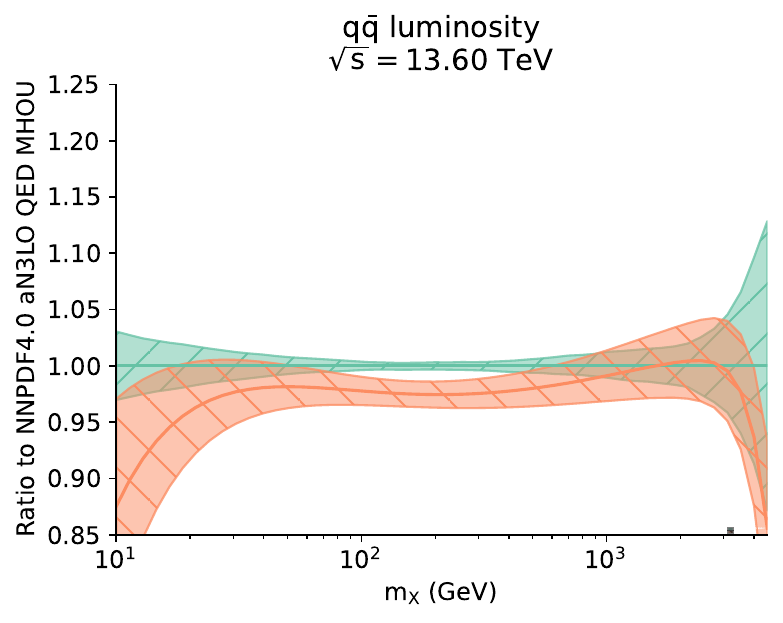}\\
\caption{The same comparison as in Fig.~\ref{fig:lumis_MSHT20} but including QED corrections.}
  \label{fig:lumis_MSHT20_QED}
\end{figure}

\begin{figure}[!p]
  \centering
  \includegraphics[width=0.45\textwidth]{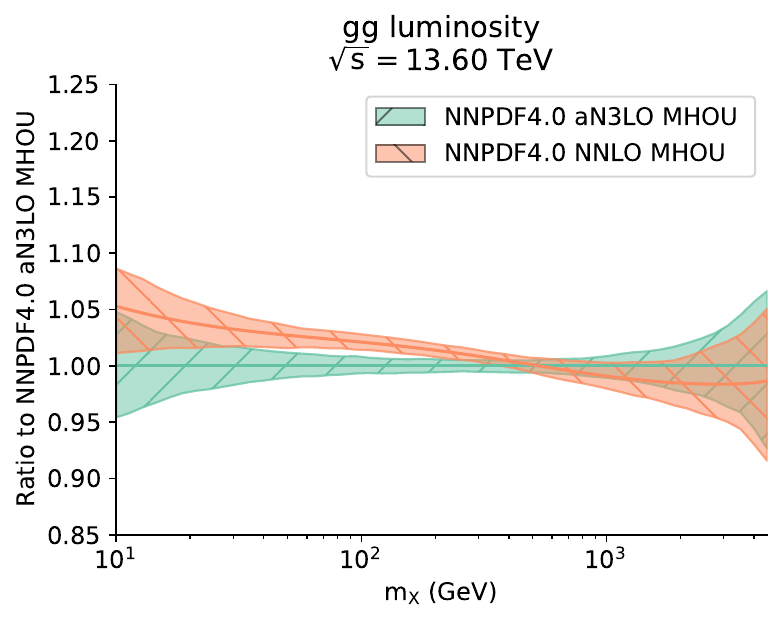}
  \includegraphics[width=0.45\textwidth]{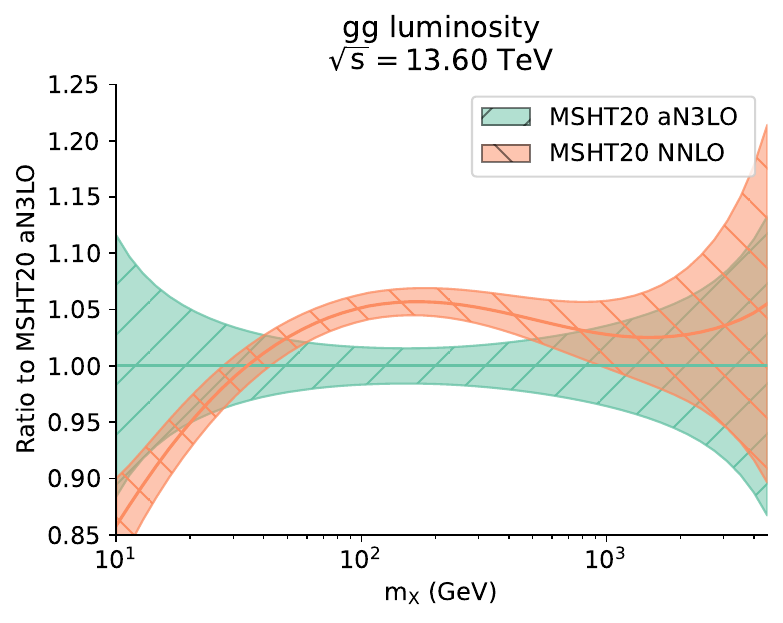}\\
  \includegraphics[width=0.45\textwidth]{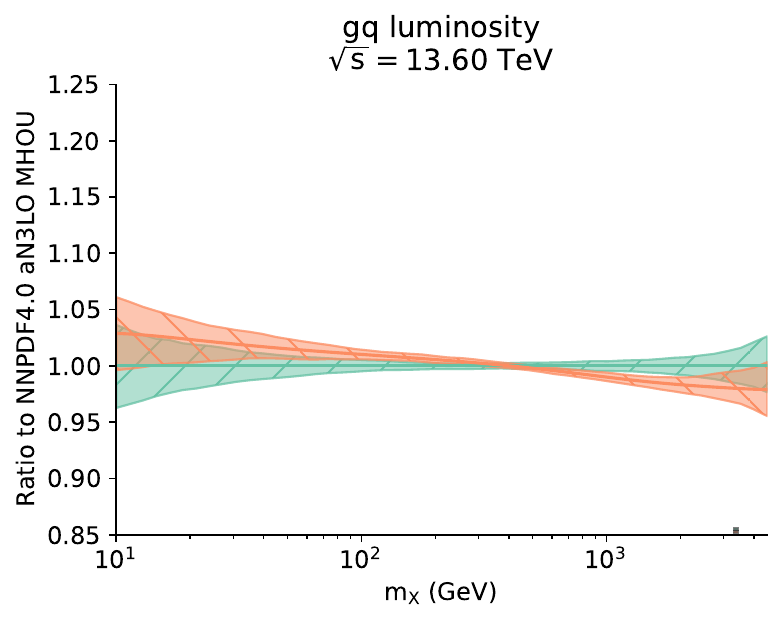}
  \includegraphics[width=0.45\textwidth]{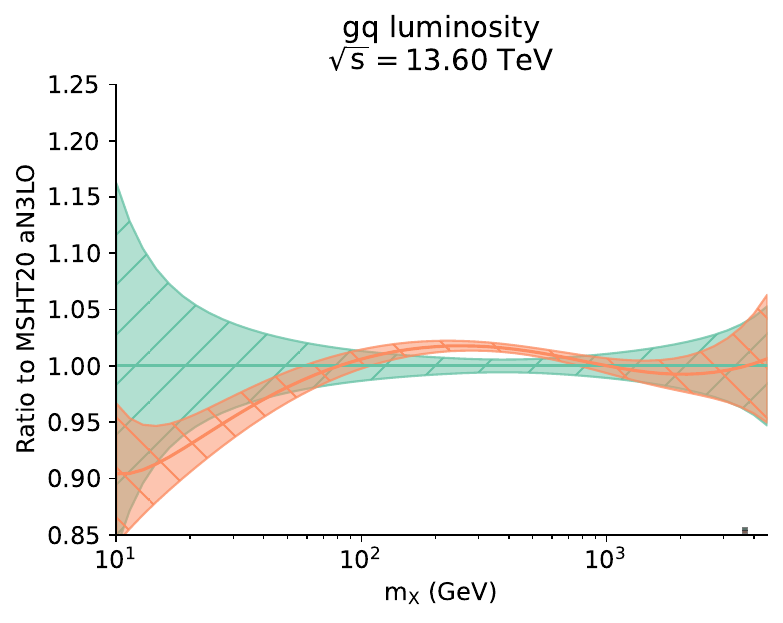}\\
  \includegraphics[width=0.45\textwidth]{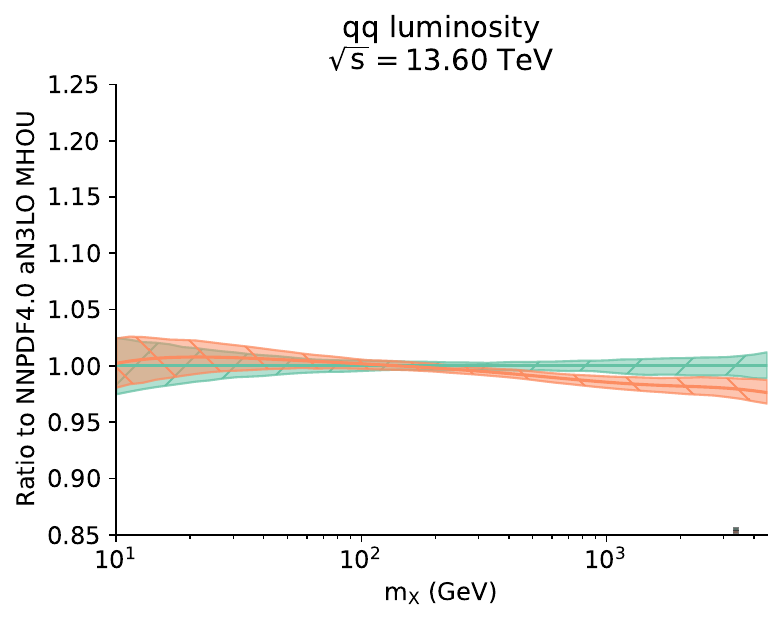}
  \includegraphics[width=0.45\textwidth]{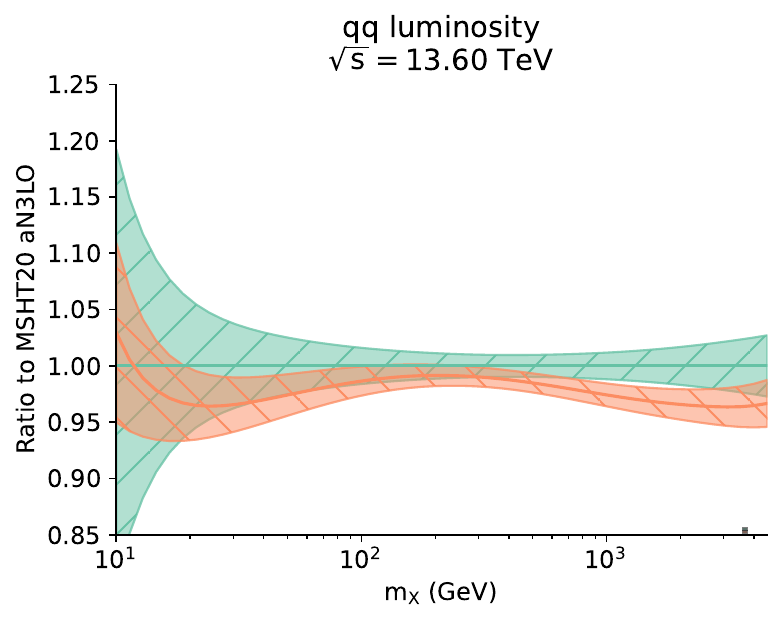}\\
  \includegraphics[width=0.45\textwidth]{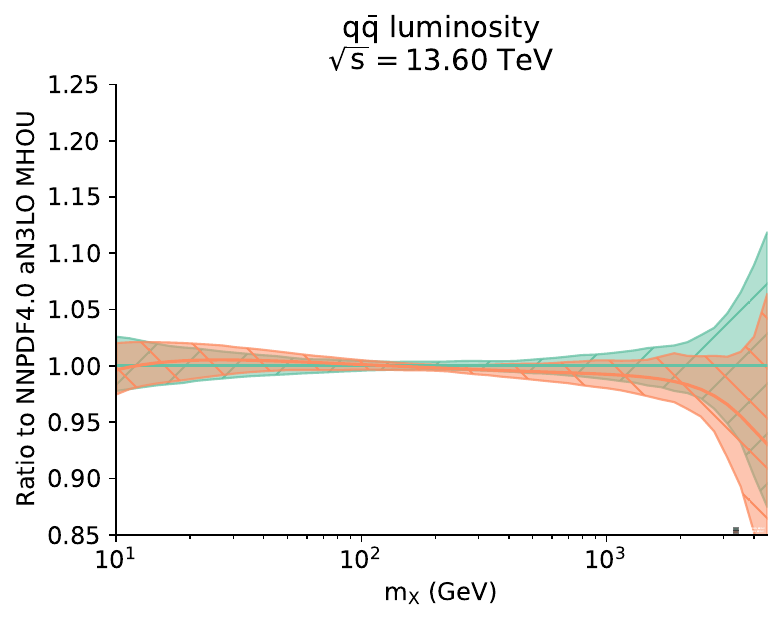}
  \includegraphics[width=0.45\textwidth]{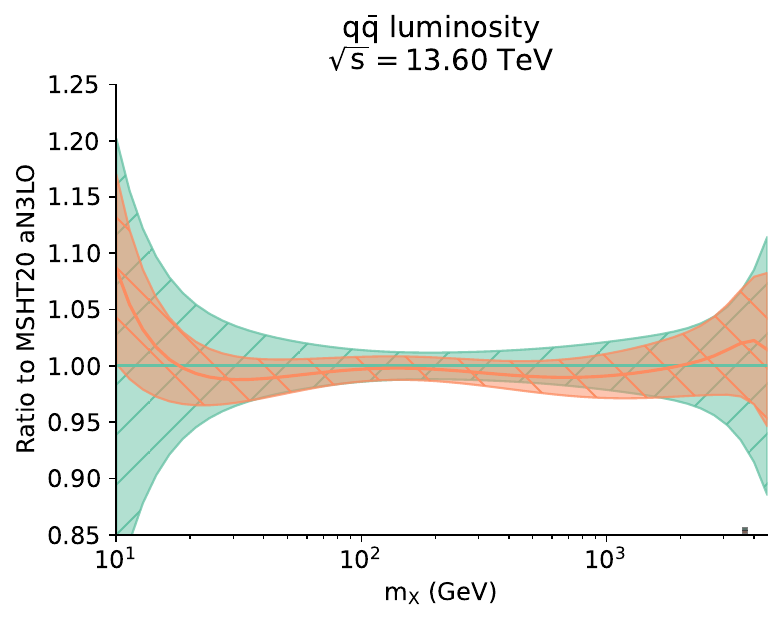}\\
  \caption{Comparison of NNLO and aN$^3$LO parton luminosities for
    NNPDF4.0 ~\cite{NNPDF:2021njg,NNPDF:2024nan}(left) and MSHT20~\cite{Bailey:2020ooq,McGowan:2022nag}
    (right), normalized to the aN$^3$LO result.}
  \label{fig:lumis_NLO_vs_N3LO}
\end{figure}

The MSHT and NNPDF aN$^3$LO parton luminosities are compared in
Figs.~\ref{fig:lumis_MSHT20}-\ref{fig:lumis_MSHT20_QED}, respectively
without and with QED effects, while, in order to appreciate better the
impact of N$^3$LO corrections, the aN$^3$LO luminosities (without QED) are compared
for each set
to the respective NNLO luminosities  in Fig.~\ref{fig:lumis_NLO_vs_N3LO}, normalized to the
aN$^3$LO result.  The pattern of differences reproduces that seen at
the level of PDFs, with a   suppression of the MSHT gluon-gluon, quark-antiquark, and
gluon-quark luminosities in comparison to NNPDF in the
$M_X\sim100$~GeV region, while the quark-quark luminosity in the two
sets is very similar. The suppression can be traced
 to the behavior of the light quark and gluon PDFs seen in
Figs.~\ref{fig:PDFs_N3LO_vs_NNLO_1}-\ref{fig:PDFs_N3LO_vs_NNLO_2} (right). Interestingly, the difference in
gluon-gluon luminosity is slightly
reduced for the QED PDF sets compared to the pure QCD sets.

The qualitative impact of the aN$^3$LO corrections on
either set for $M_X\gtrsim30$~GeV is similar (and almost identical for
the quark-quark luminosity), but with a  stronger aN$^3$LO suppression of gluon
luminosities for MSHT20.
 In particular the gluon-gluon luminosity is
suppressed for $10^2\lesssim m_X\lesssim 10^3$~GeV  by about 3\% in NNPDF4.0
and up to 6\% in MSHT20 and the gluon-quark luminosity is suppressed in the same
region by about 1\% in NNPDF4.0 and 3\% in MSHT20. Note that the impact on this comparison of
the aforementioned update studied in Ref.~\cite{Thorne:2024npj} of the splitting functions in MSHT   is 
very small due to the integration over rapidity washing out small
differences in the gluon.

These comparisons show that the differences between NNLO and aN$^3$LO
for each set are generally larger than the differences between the two
aN$^3$LO sets, or indeed the two NNLO sets, especially in the case of MSHT.
Furthermore, 
 the differences between NNLO and aN$^3$LO PDFs  are often comparable and sometimes (specifically for the  gluon-gluon luminosity) larger than the respective PDF uncertainties. These two observations taken together suggest that the
use of
aN$^3$LO PDFs is mandatory for accurate N$^3$LO phenomenology, because
using NNLO PDFs with the aN$^3$LO matrix element leads to an error
that may be larger than the PDF uncertainty, especially in the Higgs
gluon and vector boson fusion channels, that are respectively mostly
sensitive to the gluon-gluon and quark-quark luminosities.
Therefore, even though the MSHT and NNPDF aN$^3$LO PDF sets do not always agree
within uncertainties, their combination will lead to results that are
more accurate than the use of NNLO PDFs. These considerations motivate
the construction of a combined aN$3$LO PDF set.

\section{Combination of aN$^3$LO PDFs}
\label{sec:comb}

Based on the PDF comparisons presented in Sect.~\ref{sec:comp}, phenomenological predictions for hadronic
processes at N$^3$LO could be obtained by simply using the two available aN$^3$LO  PDF sets and combining results as a weighted
average. However, given that the two sets do not always agree
within uncertainties, a more conservative way of estimating the
uncertainty on the final result is advisable, such as the
so-called PDG prescription and variations
thereof~\cite{Erler:2020bif}, see also Sect.~12.5 of Ref.~\cite{LHCHiggsCrossSectionWorkingGroup:2011wcg}.

An effective way of arriving at a conservative uncertainty estimate in
our case is to produce unweighted Monte Carlo
combined PDF sets~\cite{Butterworth:2015oua,Carrazza:2015hva}. This is done by first turning all the PDF sets to be combined into a set of Monte Carlo replicas, which for Hessian sets
can be done using the methodology of Ref.~\cite{Watt:2012tq}, and then
merging an equal number of replicas from each set into a single
replica set.  The merged set then corresponds to a probability
distribution that is the equally likely combination of the probability
distribution of the input sets, and thus will lead to uncertainties that
encompass those of the sets that are being combined. This way if combining uncertainties leads to results that are more conservative than those that would be obtained by performing a weighted average, or possibly including correlations from the underlying data; however, correlations cannot~\cite{Ball:2021dab} be reliably taken into account, and differences due to different but equally plausible underlying methodological assumptions are thereby correctly accounted for.

We have produced this combination by merging 100 Monte Carlo replicas
produced using the method of Ref.~\cite{Watt:2012tq} from the default
Hessian aN$^3$LO set of Ref.~\cite{McGowan:2022nag} after
symmetrization of the 
Hessian uncertainties, with 100 replicas from the MHOU set of Ref.~\cite{NNPDF:2024dpb}. In order to improve numerical stability of results, for the NNPDF4.0 PDFs the $(x,Q^2)$ interpolation {\sc\small LHAPDF} grid has been recast to match the MSHT20 grid. 
We have then repeated the procedure for the QED variants presented in
Refs.~\cite{Cridge:2023ryv,Barontini:2024dyb}. We will henceforth
denote these combined PDF sets as  MSHT20xNNPDF40\_aN3LO; the
corresponding {\sc\small LHAPDF} grid files are {\tt  MSHT20xNNPDF40\_an3lo} and 
{\tt  MSHT20xNNPDF40\_an3lo\_qed}.
We have also produced combined pure QCD and QCD$\otimes$QED  NNLO sets obtained
using the same procedure, but 
now starting from the NNPDF4.0 and MSHT  NNLO PDF sets. The sole
purpose of these sets is to provide a baseline to the corresponding
aN$^3$LO sets, in order to assess the effect of N$^3$LO corrections
with everything else unchanged. In particular, they should not be
viewed as a substitute for the PDF4LHC21 NNLO combined sets. 
These sets are
denoted as  MSHT20xNNPDF40\_NNLO, and the
corresponding {\sc\small LHAPDF} grid files
{\tt  MSHT20xNNPDF40\_nnlo} and {\tt  MSHT20xNNPDF40\_nnlo\_qed}. Note that, as discussed in Sect.~\ref{sec:comp},
the aN$^3$LO sets do include while the NNLO sets do not include
uncertainties due to MHO in the theory predictions used for PDF
determination.

It should always be kept in mind that, as already mentioned,
unlike the PDF4LHC21 combined
PDFs~\cite{PDF4LHCWorkingGroup:2022cjn}, these  aN$^3$LO and NNLO
MSHT20xNNPDF40 PDFs are an
unmodified combination of the two input sets, which are based on
somewhat different theoretical and methodological
assumptions. They should be 
viewed as a means to obtain accurate but conservative N$^3$LO
predictions for LHC 
processes, not as an optimized PDF combined set. For example, due to
the aforementioned different treatment of  heavy quarks, the combined PDFs
are only reliable for scales well above the heavy quark thresholds, and
should not be used in regions where heavy quark mass effects are
relevant, such as Higgs production via bottom quark fusion.
However, we will show below that for scales $M_X\gtrsim 50$~GeV  the
effects of these 
different treatments of quark 
masses are in general smaller than the differences between 
NNLO vs.  aN$^3$LO PDFs. 

\begin{figure}[!p]
  \centering
  \includegraphics[width=0.45\textwidth]{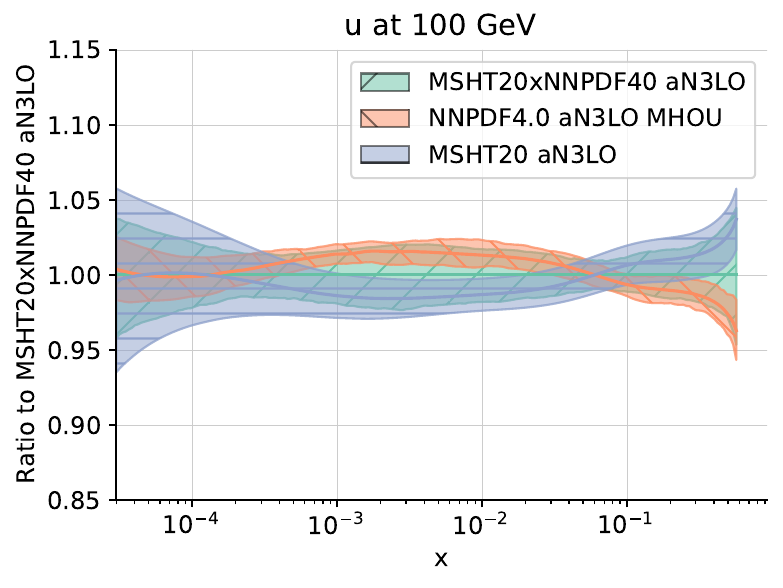}
  \includegraphics[width=0.45\textwidth]{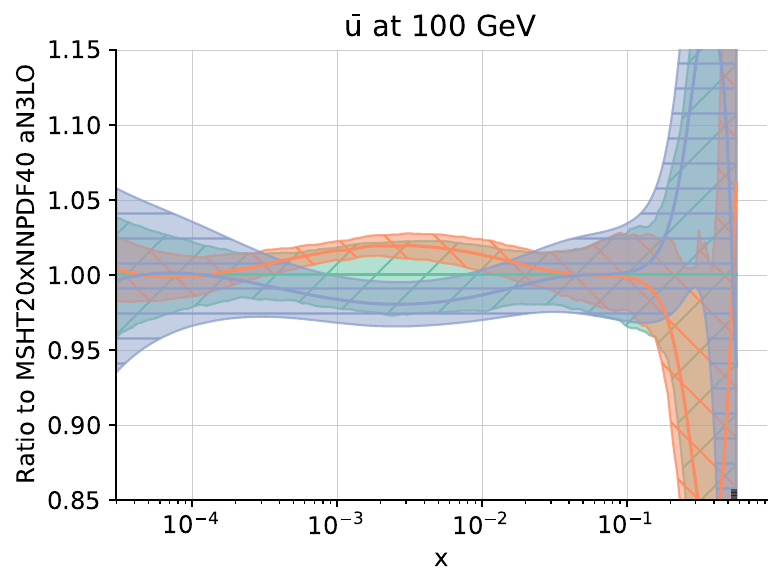}\\
  \includegraphics[width=0.45\textwidth]{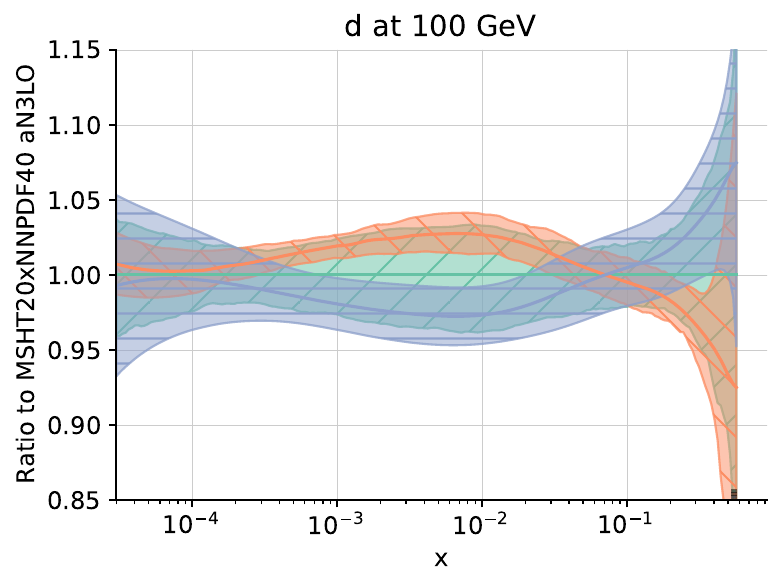}
  \includegraphics[width=0.45\textwidth]{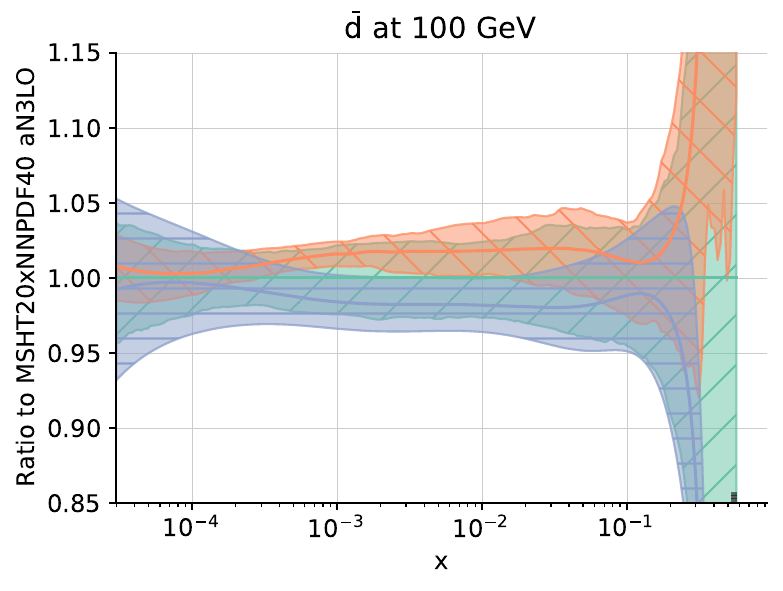}\\
  \includegraphics[width=0.45\textwidth]{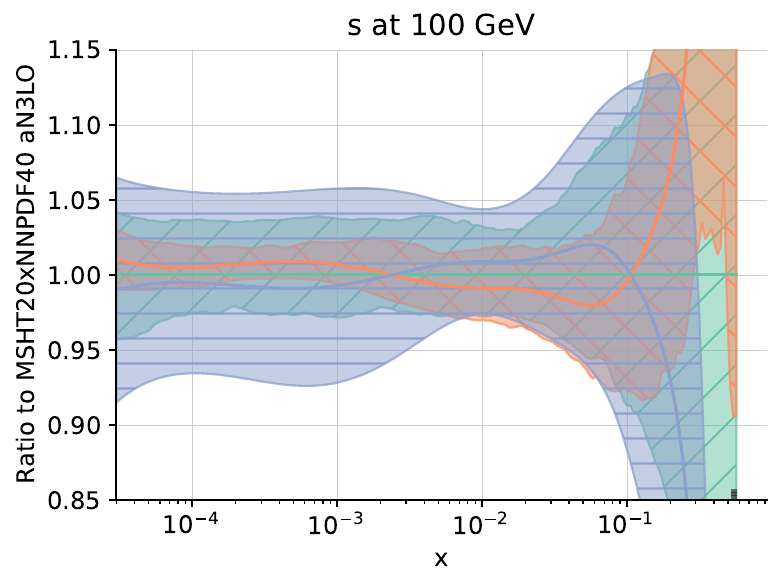}
  \includegraphics[width=0.45\textwidth]{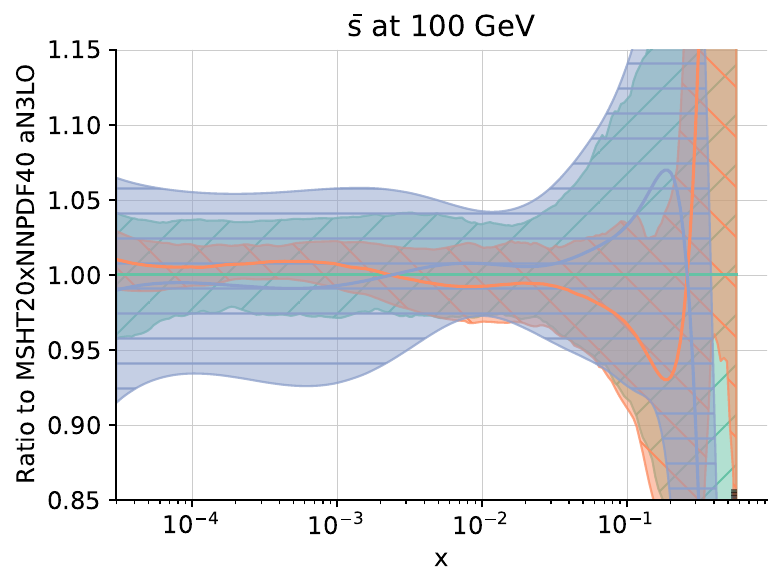}\\
  \includegraphics[width=0.45\textwidth]{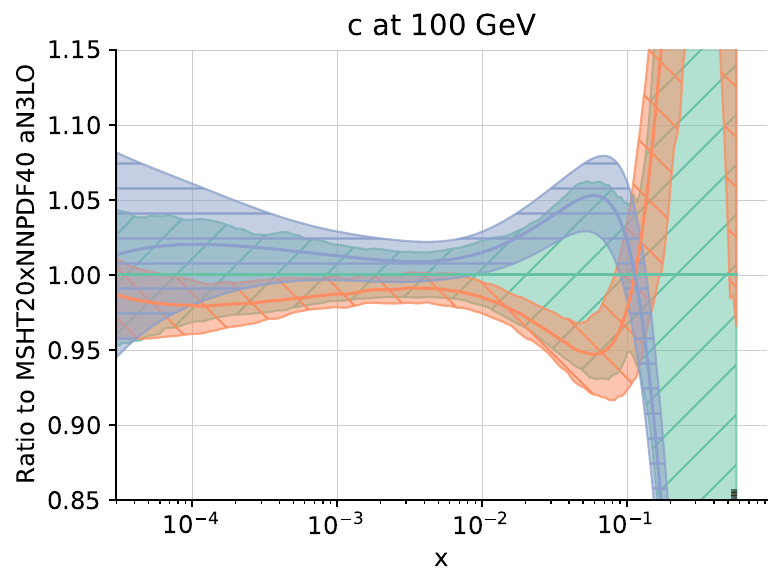}
  \includegraphics[width=0.45\textwidth]{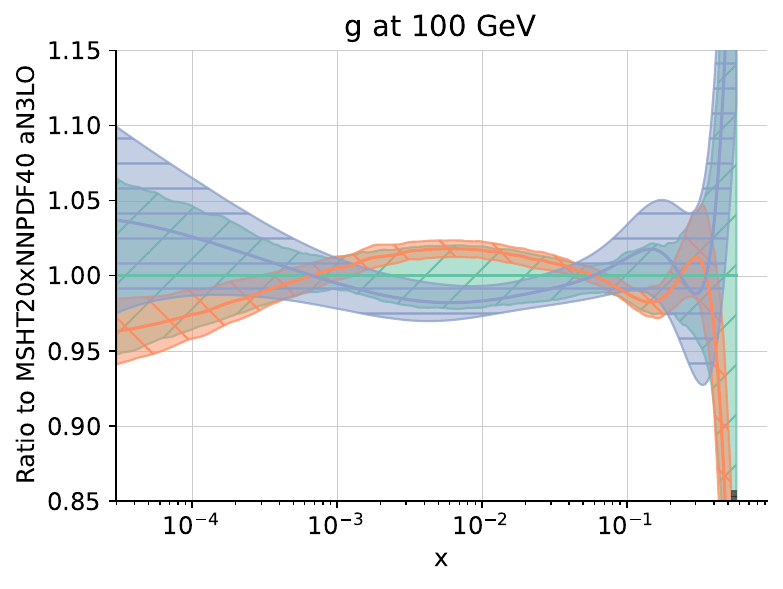}\\
  \caption{The combined  MSHT20xNNPDF40\_aN3LO PDFs, compared to the
    NNPDF and MSHT aN$^3$LO PDFs that enter the combination~\cite{NNPDF:2024nan,McGowan:2022nag}, shown as
    a ratio to  MSHT20xNNPDF40\_aN3LO.}  \label{fig:PDF_combination_QCD}
\end{figure}

\begin{figure}[!p]
  \centering
  \includegraphics[width=0.45\textwidth]{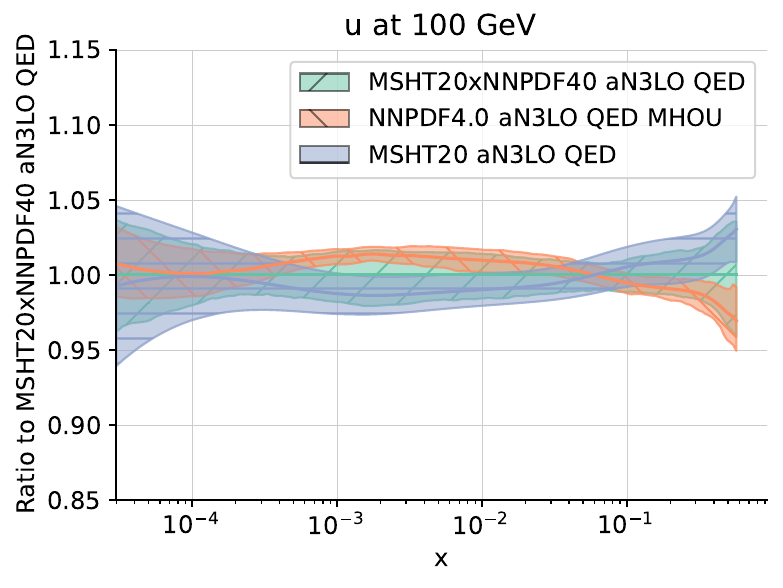}
  \includegraphics[width=0.45\textwidth]{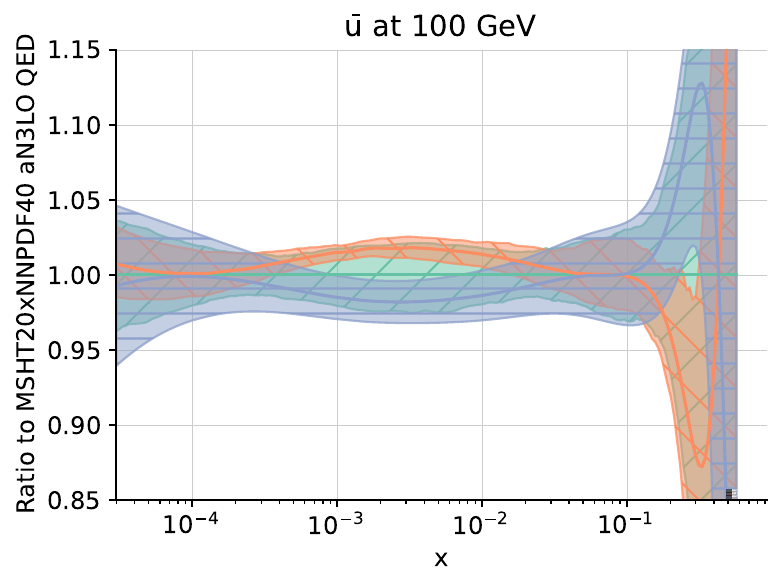}\\
  \includegraphics[width=0.45\textwidth]{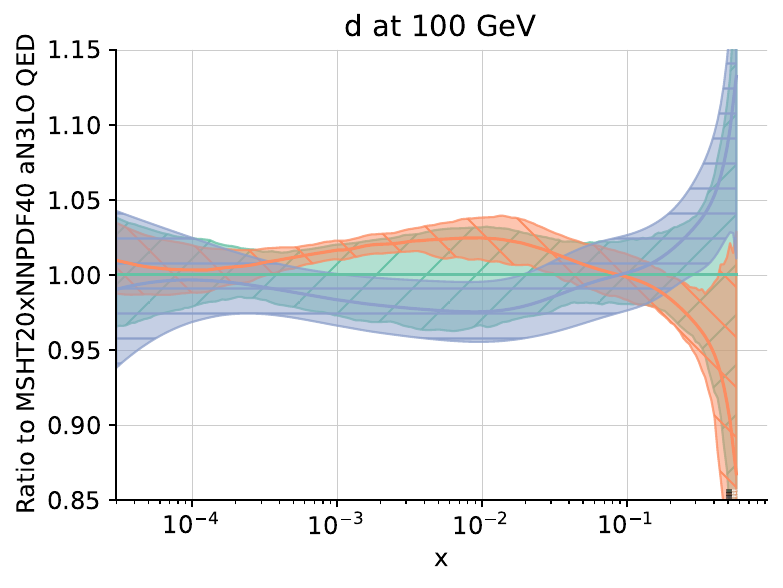}
  \includegraphics[width=0.45\textwidth]{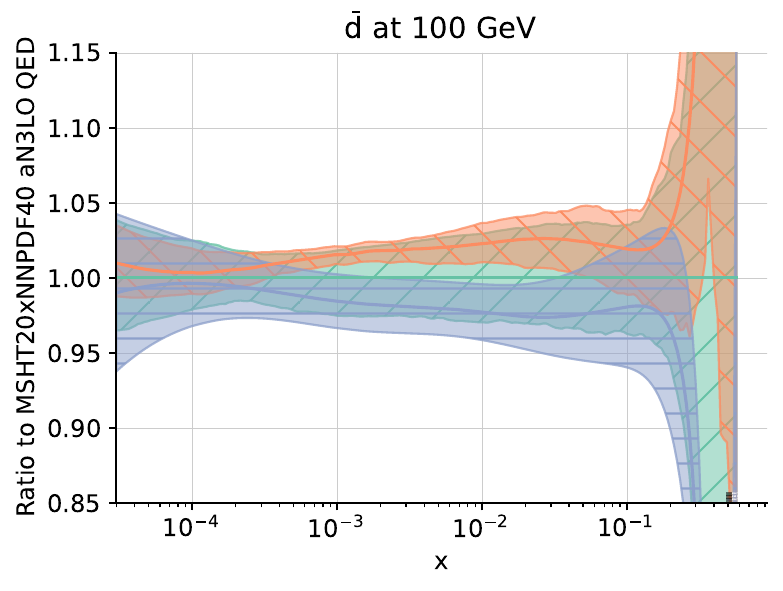}\\
  \includegraphics[width=0.45\textwidth]{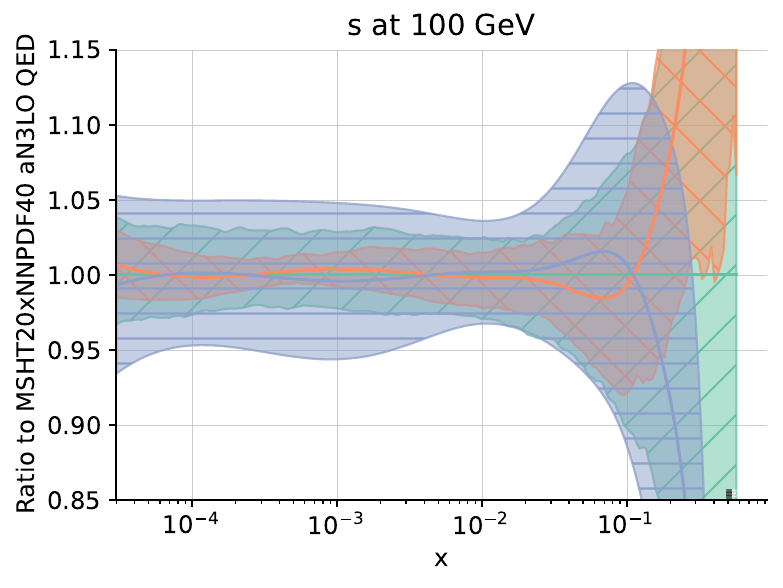}
  \includegraphics[width=0.45\textwidth]{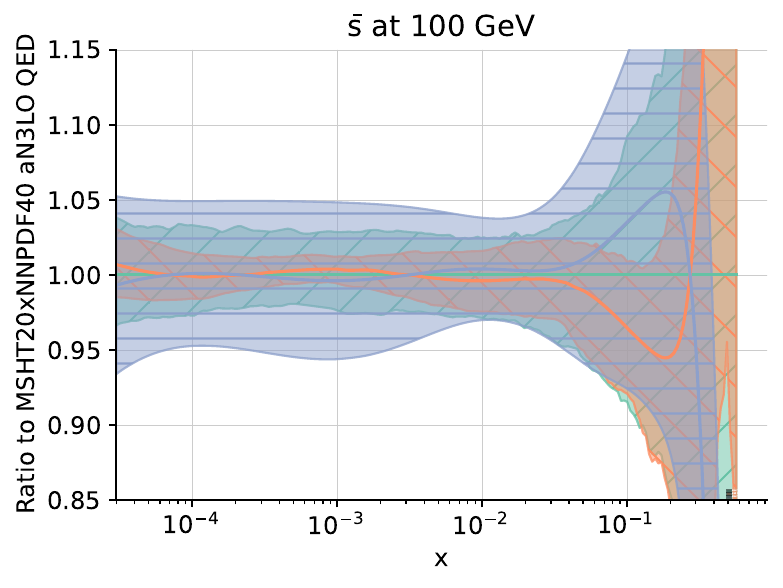}\\
  \includegraphics[width=0.45\textwidth]{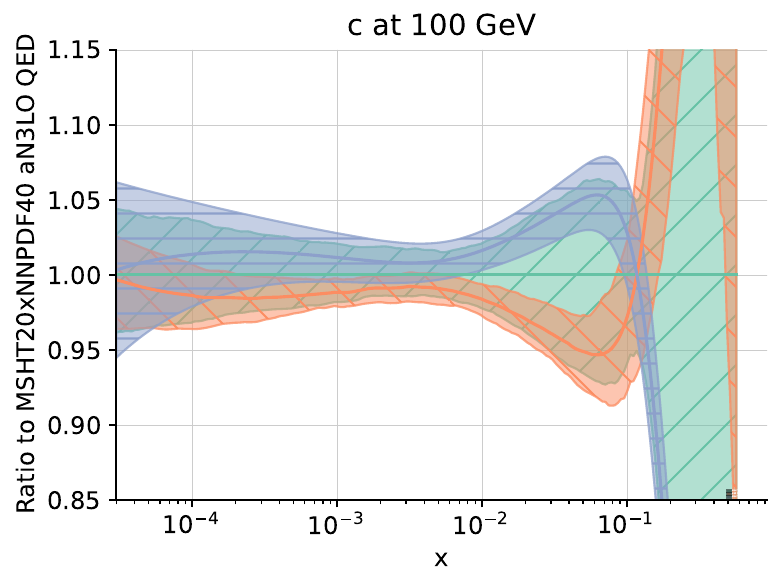}
  \includegraphics[width=0.45\textwidth]{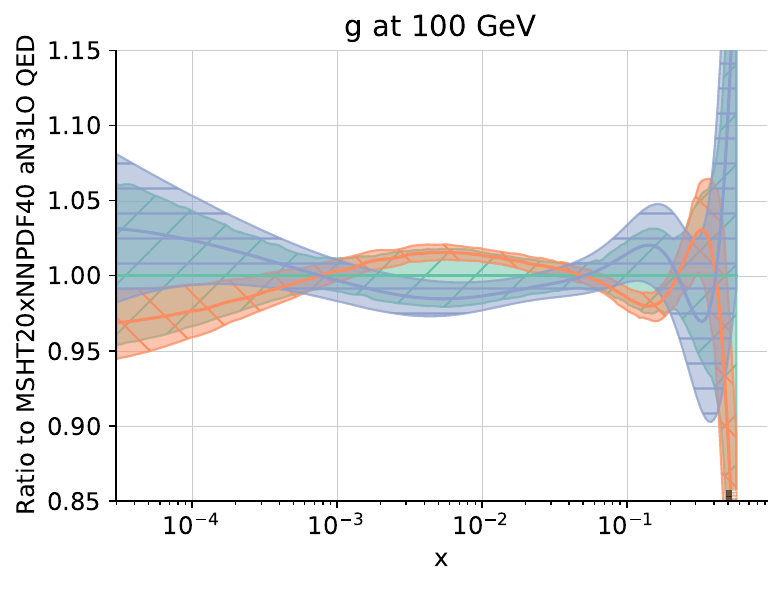}\\
  \caption{Same as Fig.~\ref{fig:PDF_combination_QCD}, but for the QED
    PDF sets.}
  \label{fig:PDF_combination_QED}
\end{figure}

The combined aN$^3$LO  PDFs are displayed  at $Q=100$~GeV in
Figs.~\ref{fig:PDF_combination_QCD}-\ref{fig:PDF_combination_QED},
respectively without and with QED effects, together with the two sets
that are being merged, shown as a ratio to the
central value of the combined set. The central value of the combined
set is, by construction, the unweighted average of the two starting
sets, and the uncertainty is seen to encompass that of the two sets
that enter the combination.

\begin{figure}[!t]
  \centering
  \includegraphics[width=0.45\textwidth]{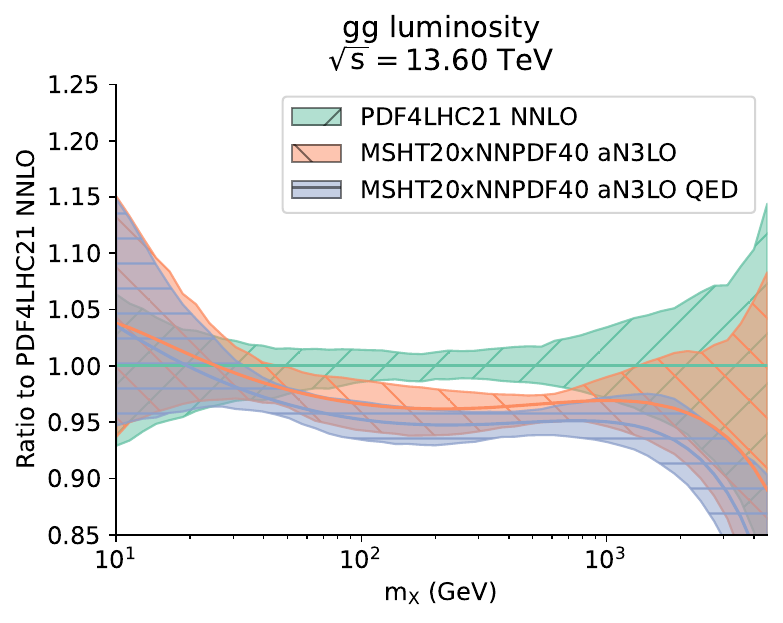}
  \includegraphics[width=0.45\textwidth]{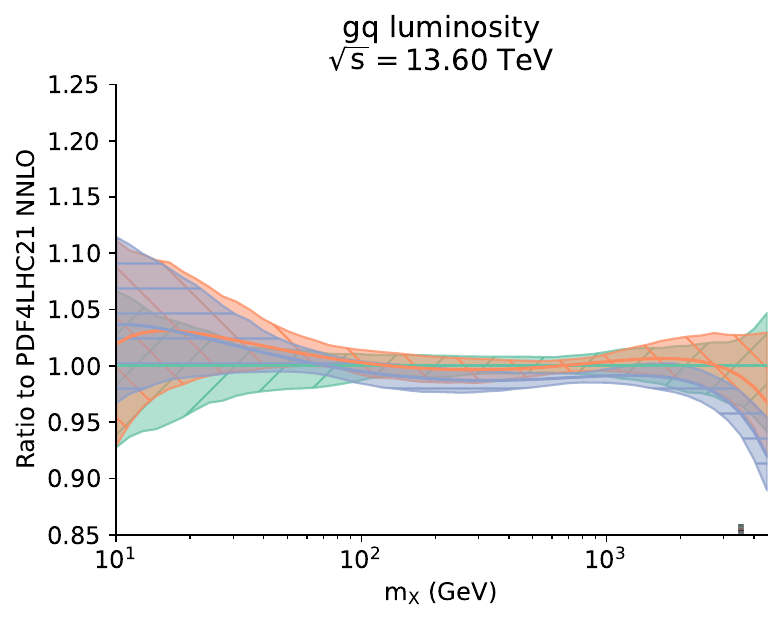}\\
  \includegraphics[width=0.45\textwidth]{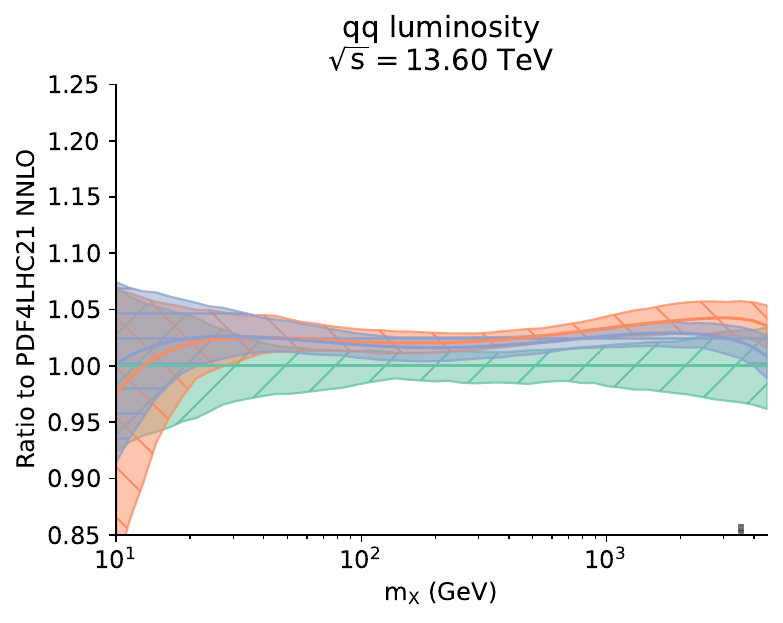}
  \includegraphics[width=0.45\textwidth]{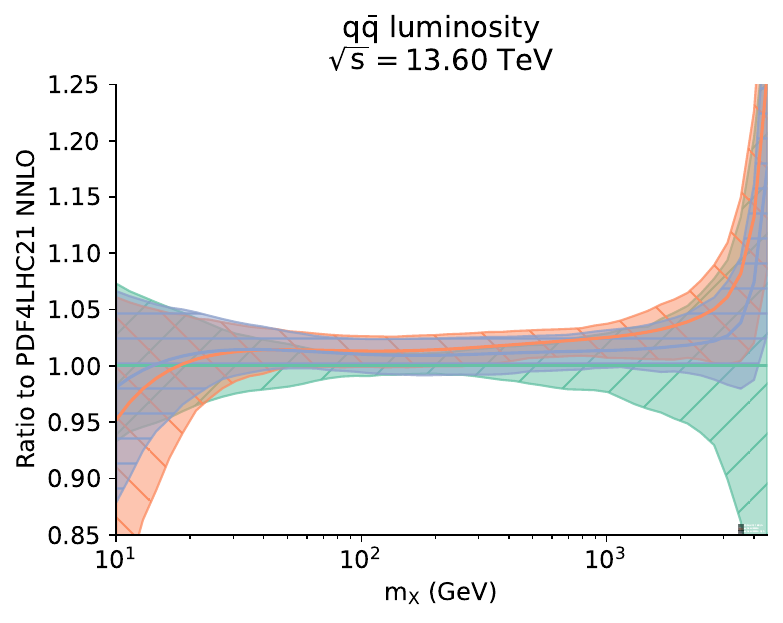}\\
  \caption{ Parton luminosities for the LHC  $\sqrt{s}=13.6$~TeV
    computed from  the  MSHT20xNNPDF40\_aN3LO pure QCD and QCD+QED sets, compared to the NNLO
PDF4LHC21 result~\cite{PDF4LHCWorkingGroup:2022cjn}, and shown a ratio to the latter.}
  \label{fig:lumi-n3lo-PDF4HXSWG}
\end{figure}

\begin{figure}[!t]
  \centering
  \includegraphics[width=0.45\textwidth]{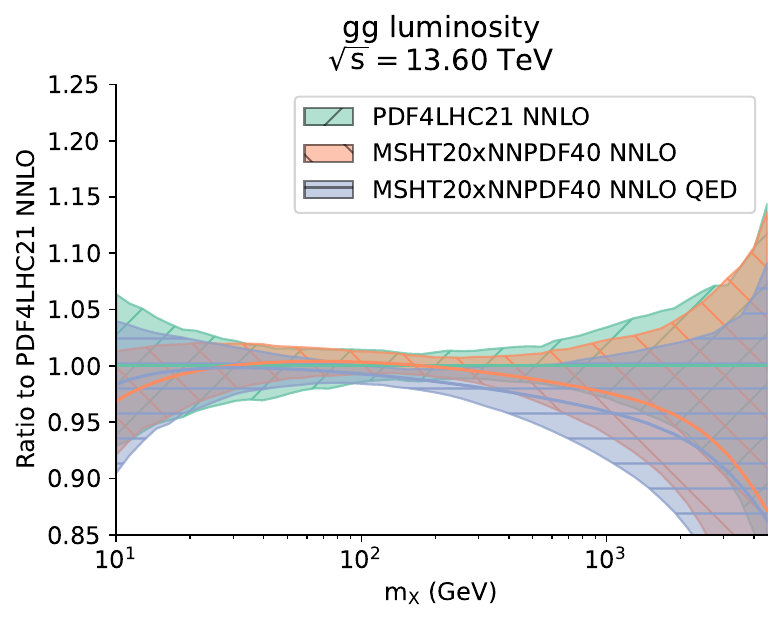}
  \includegraphics[width=0.45\textwidth]{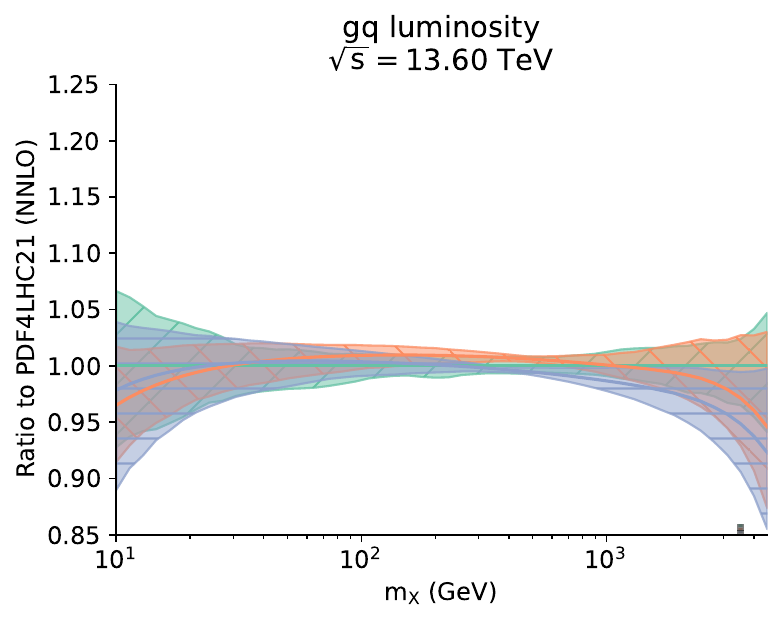}\\
  \includegraphics[width=0.45\textwidth]{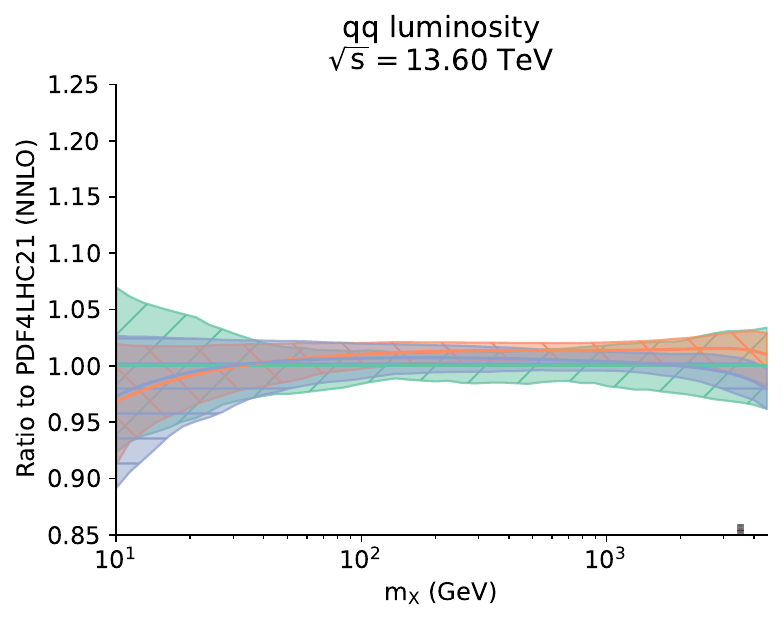}
  \includegraphics[width=0.45\textwidth]{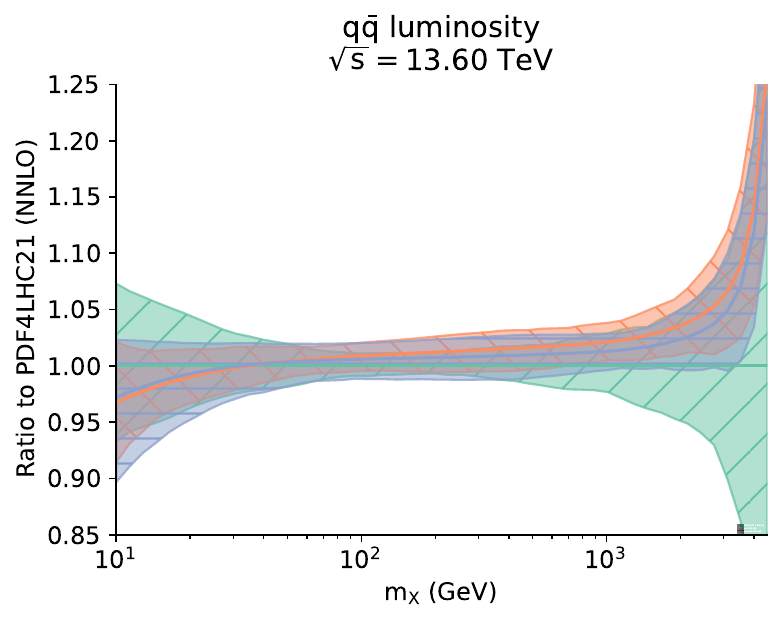}\\
  \caption{Comparison of NNLO parton luminosities for the LHC with  $\sqrt{s}=13.6$~TeV: the 
      MSHT20xNNPDF40\_nnlo pure QCD and QCD+QED results are compared to the
     PDF4LHC21 combination~\cite{PDF4LHCWorkingGroup:2022cjn}.}
  \label{fig:lumi-nnlo-PDF4HXSWG}
\end{figure}

In order to assess the  impact of the N$^3$LO corrections, in
Fig.~\ref{fig:lumi-n3lo-PDF4HXSWG} we compare luminosities for the LHC
$\sqrt{s}=13.6$~TeV at NNLO and
 aN$^3$LO, in the latter case both without and then with QED corrections,
 shown as a ratio to the NNLO result, computed using the PDF4LHC21
 combined NNLO PDF set~\cite{PDF4LHCWorkingGroup:2022cjn}. 
For reference, the NNLO sets are compared to each other  in
Fig.~\ref{fig:lumi-nnlo-PDF4HXSWG}, where we show the two different
combinations, PDF4LHC21 and  MSHT20xNNPDF40\_NNLO, the latter both
without and with QED corrections.
It is clear that for the gluon-gluon and quark-quark luminosities
at all scales  $M_X\gtrsim100$~GeV the
 impact of the aN$^3$LO corrections is quite significant and in the
 gluon-gluon channel well outside the NNLO uncertainty band. On the
 other hand, the difference between the two NNLO combinations is
 significantly smaller, with each of them well within the uncertainty
 band of the other, thereby showing that the significance of the aN$^3$LO does not depend on the NNLO baseline that one compares it to. 
  For the
 gluon-gluon luminosity the impact of QED corrections is also
 significant, both at NNLO and aN$^3$LO.

 It is interesting to observe
 that for the gluon--gluon luminosity in the range
 $30\lesssim M_X\lesssim300$~GeV the difference between aN$^3$LO and
 NNLO is significantly larger than the difference between NNLO and
 NLO, with the effect peaking around $M_X\sim100$~GeV (see Fig.~24 of
 Ref.~\cite{NNPDF:2024nan})), though somewhat reduced upon the
 inclusion of MHOUs. This can be understood as the consequence of the
fact that at N$^3$LO there is a leading small $x$ logarithmic contribution to
$P_{gg}$ and $P_{gq}$, which instead accidentally vanishes both at NLO and
NNLO. This results in N$^3$LO corrections leading to sizable contribution to gluon evolution even
at intermediate $x$: indeed, both splitting functions at  N$^3$LO are
qualitatively similar to the form that one obtains when including into
the splitting function full small-$x$ resummation~\cite{NNPDF:2024nan}. Furthermore, 
the effect of N$^3$LO corrections on splitting functions, heavy flavour
contributions, and light quark coefficient functions all conspire to push the
gluon in the same direction for all $x\lesssim0.1$ (see Fig.~44 of Ref.~\cite{McGowan:2022nag}).

\section{Implications for Higgs production}
\label{sec:higgs}

\begin{figure}[p!]
  \centering
  \includegraphics[width=0.49\textwidth]{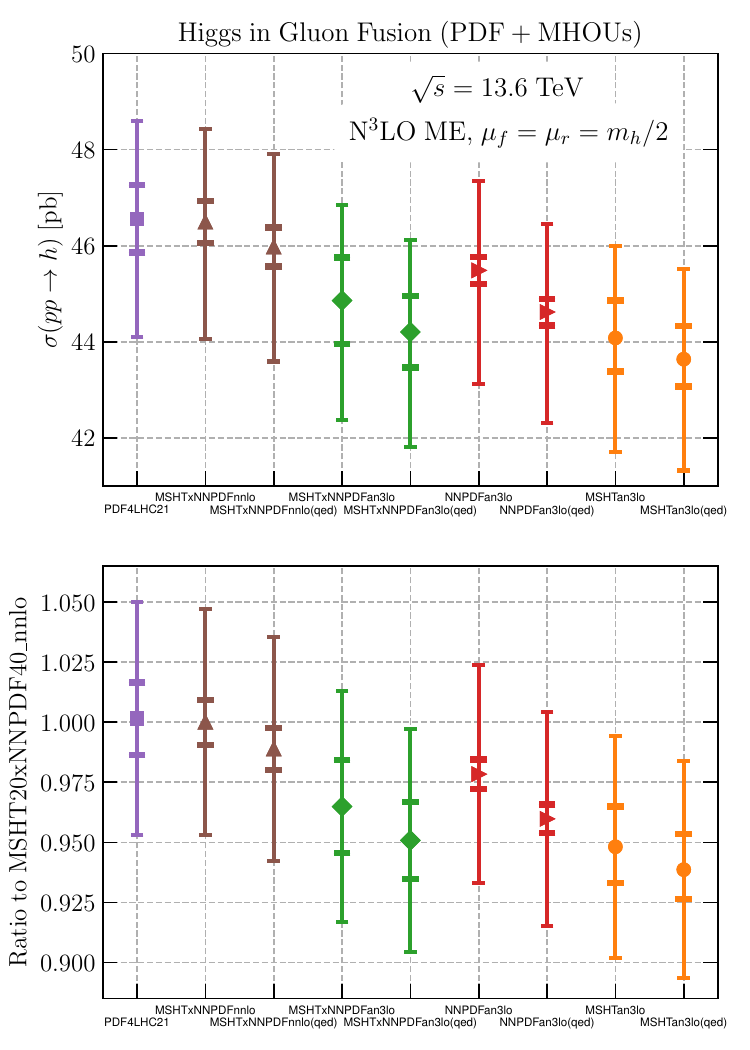}
\includegraphics[width=0.49\textwidth]{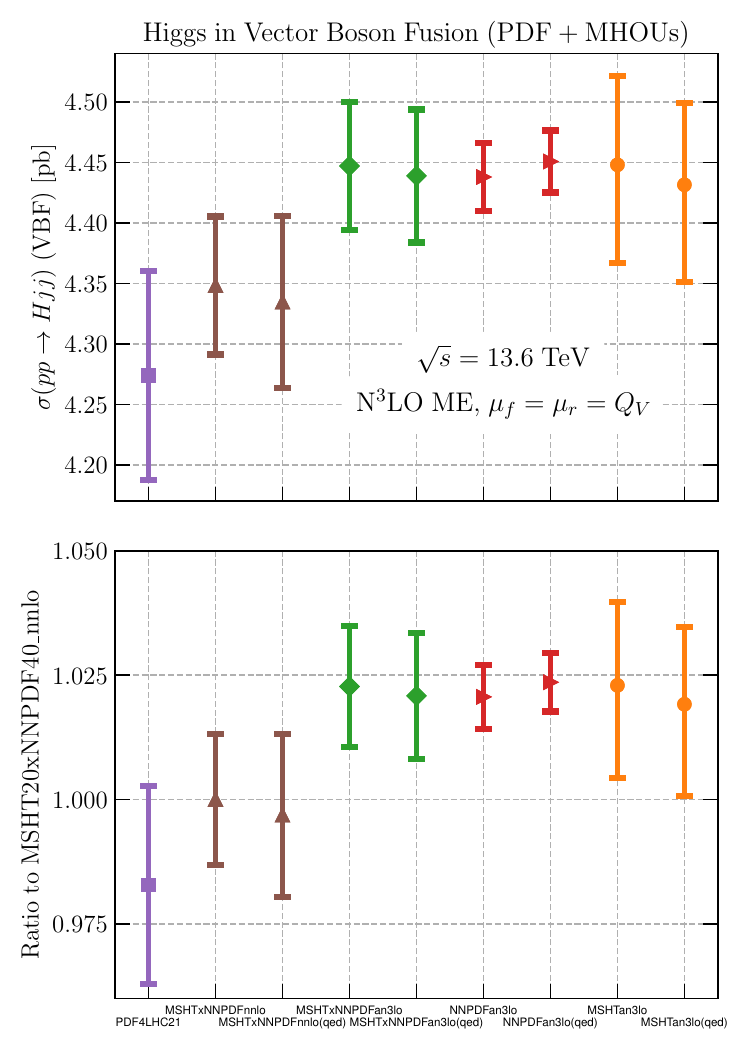}   \includegraphics[width=0.49\textwidth]{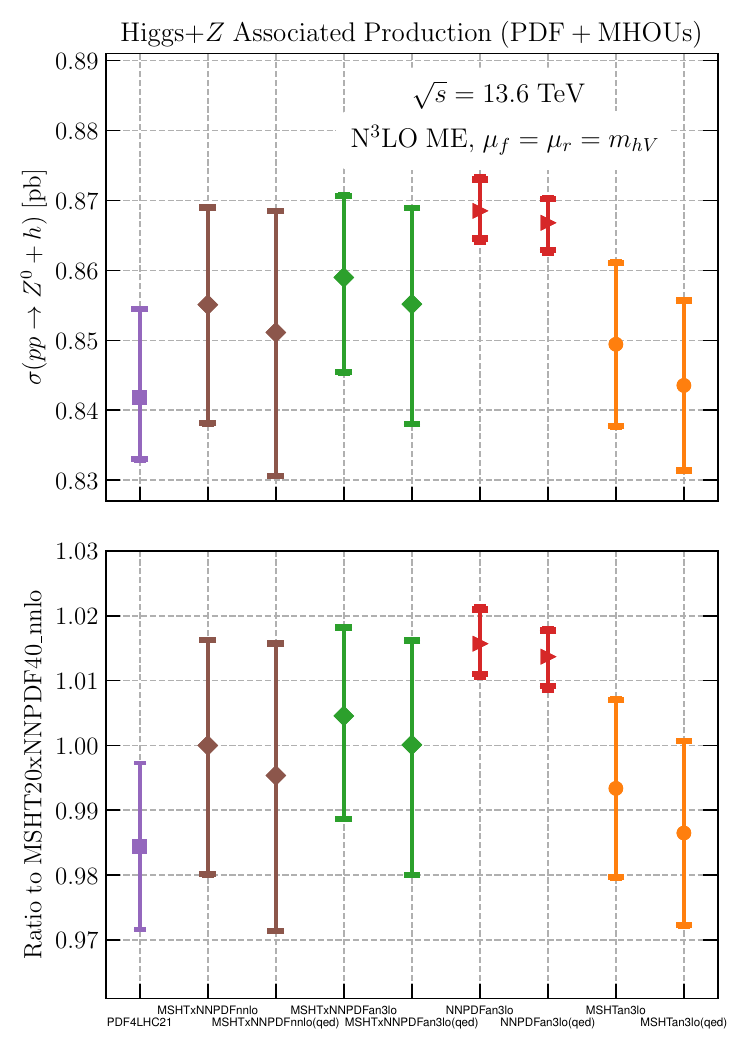}
\includegraphics[width=0.49\textwidth]{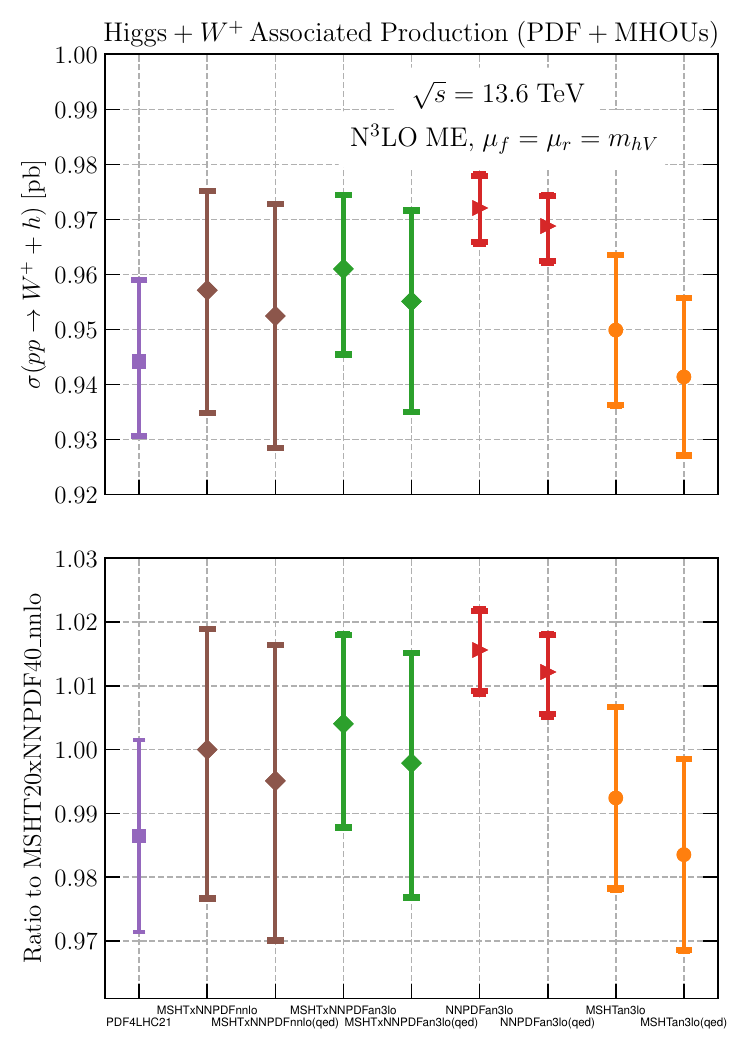}
  \caption{The cross-sections of Tables~\ref{table:pheno}-\ref{table:pheno2}, shown both in absolute scale (top) or as ratios to the result found using the {\tt MSHT20xNNPDF40\_nnlo} baseline combination.
  In all cases, N$^3$LO matrix elements are used.
    The results for $hW^-$ are qualitatively similar as those for $hW^+$ and not shown.
    The inner interval is the pure PDF uncertainty (first uncertainty in Tables~\ref{table:pheno}-\ref{table:pheno2}) while the outer interval is the sum in quadrature of the PDF uncertainty and the MHOU uncertainty on the hard cross section (second uncertainty in Tables~\ref{table:pheno}-\ref{table:pheno2}). 
}
  \label{fig:higgs-ggF-n3lo-HXSWG}
\end{figure}

We now present predictions for Higgs production at the LHC in gluon fusion, in vector boson fusion (VBF), and in association with a weak gauge boson ($hV$).
Results are collected in Tables~\ref{table:pheno}-\ref{table:pheno2} and displayed in
Fig.~\ref{fig:higgs-ggF-n3lo-HXSWG}.
All results shown are computed  for the LHC with $\sqrt{s}=13.6$~TeV using
N$^3$LO matrix elements, and  only differ in the input PDF set. 
No QED or electroweak corrections are included in the matrix
element, and in particular  photon--initiated processes are not
included. The impact of these on the processes considered here is  assessed e.g. in Ref.~\cite{NNPDF:2024djq}, it is negligible for gluon fusion, of order of a percent for VBF and a few percent for associate production, and it is similar for all PDF sets considered here as the photon PDF in the NNPDF4.0 and MSHT20 sets is quite similar, so it does not affect the comparisons presented here.

The  cross-sections are computed using the {\tt ggHiggs} code~\cite{Bonvini:2014jma,Bonvini:2013kba} for gluon fusion, {\tt n3loxs}~\cite{Baglio:2022wzu} for associated production, and {\tt proVBFH}~\cite{Dreyer:2016oyx,Dreyer:2018qbw} for VBF.
The central factorization and renormalization scales are set to $\mu_F=\mu_R=m_H/2$ for Higgs production in gluon fusion, $\mu_F=\mu_R=Q_{V}$ (the vector boson virtuality) for VBF, and $\mu_F=\mu_R=m_{hV}$, the event-by-event invariant mass of the $hV$ system, in associated production. 
In Fig.~\ref{fig:higgs-ggF-n3lo-HXSWG} we do not display the $hW^-$ cross-sections since they have the same qualitative behavior as the $hW^+$ cross-sections, as can also be observed from Table~\ref{table:pheno}.

The uncertainties shown in Tables~\ref{table:pheno}-\ref{table:pheno2} and in
Fig.~\ref{fig:higgs-ggF-n3lo-HXSWG} are the pure PDF uncertainty
(first uncertainty in the table and inner error bar in the figure) as
well as the sum in quadrature of the PDF uncertainty and MHO
uncertainties on the matrix element, the latter evaluated with the 7
point scale variation prescription (second uncertainty in the table
and outer error bar in the figure). 
For the VBF and $hV$ production channels the two uncertainties
essentially coincide, because scale variations are negligible and  the
PDF uncertainty dominates, while in the gluon-fusion cross-sections scale errors represent the largest theoretical uncertainty.

\begin{table}[t!]
  \centering
  \small
   \renewcommand{\arraystretch}{2.0}
   \begin{tabularx}{\textwidth}
   {|X|l|l|l|}
     \toprule
  PDF set  
  &
  pert. order (PDF)
  &
  $\sigma(gg\to h)$  
       &  
       $\sigma(h~{\rm VBF})$\\
  \midrule
      {\tt PDF4LHC21\_mc}  & NNLO$_{\rm QCD}$  
      & 
      $46.56^{+1.5\%}_{-1.5\%}\,^{+ 4.4\%}_{-5.3\%}$   
      & $4.27^{+2.0\%}_{-2.0\%}\,^{+2.0\%}_{-2.0\%}$
      \\
       {\tt  MSHT20xNNPDF40\_nnlo}  & NNLO$_{\rm QCD}$ &
       $46.49^{+0.9\%}_{-0.9\%}\,^{+ 4.2\%}_{-5.3\%}$   
      & 
      $4.35^{+1.3\%}_{-1.3\%}\,^{+1.3\%}_{-1.3\%}$
      \\
       {\tt  MSHT20xNNPDF40\_nnlo\_qed}  & NNLO$_{\rm QCD}\otimes {\rm NLO}_{\rm QED}$ 
       & 
       $45.97^{+0.9\%}_{-0.9\%}\,^{+4.3\%}_{-5.4\%}$  
      & 
      $4.34^{+1.6\%}_{-1.6\%}\,^{+1.6\%}_{-1.6\%}$
      \\
      \midrule
       {\tt  MSHT20xNNPDF40\_an3lo}  
      &
      aN$^3$LO$_{\rm QCD}$
      & 
      $44.86^{+2.0\%}_{-2.0\%}\,^{+ 4.4\%}_{-5.6\%}$
      &
    $4.45^{+1.2\%}_{-1.2\%}\,^{+1.2\%}_{-1.2\%}$\\
      {\tt MSHT20xNNPDF40\_an3lo\_qed} 
      &
       aN$^3$LO$_{\rm QCD}\otimes {\rm NLO}_{\rm QED}$
      &  
      $44.20^{+ 1.7\%}_{-1.7\%}\,^{+ 4.3\%}_{-5.4\%}$
      & 
    $4.44^{+1.2\%}_{-1.2\%}\,^{+1.2\%}_{-1.2\%}$\\ 
\midrule
      {\tt NNPDF40\_an3lo\_as\_01180\_mhou} & 
       aN$^3$LO$_{\rm QCD}$
      & 
      $45.49^{+0.6\%}_{-0.6\%}\,^{+ 4.1\%}_{-5.2\%}$
      &
    $4.44^{+0.6\%}_{-0.6\%}\,^{+0.7\%}_{-0.6\%}$\\
      {\tt NNPDF40\_an3lo\_as\_01180\_qed\_mhou} 
      & 
       aN$^3$LO$_{\rm QCD}\otimes {\rm NLO}_{\rm QED}$
      &  
      $44.62^{+0.6\%}_{-0.6\%}\,^{+ 4.1\%}_{-5.2\%}$
      &
    $4.45^{+0.6\%}_{-0.6\%}\,^{+0.6\%}_{-0.6\%}$\\ 
    \midrule
      {\tt MSHT20an3lo\_as118}  
      &
       aN$^3$LO$_{\rm QCD}$
      & 
      $44.08^{+1.8\%}_{-1.6\%}\,^{+ 4.4\%}_{-5.6\%}$ 
      & 
$4.44^{+1.6\%}_{-1.8\%}\,^{+1.7\%}_{-1.8\%}$
      \\
    {\tt MSHT20qed\_an3lo} 
    & 
     aN$^3$LO$_{\rm QCD}\otimes {\rm NLO}_{\rm QED}$
    &
    $43.63^{+1.6\%}_{-1.3\%}\,^{+ 4.3\%}_{-5.3\%}$ 
    &
$4.43^{+1.5\%}_{-1.8\%}\,^{+1.5\%}_{-1.8\%}$ \\
     \midrule
    \midrule
    $ \Delta^{\rm exact}_{\rm NNLO} $ (NNPDF4.0)   
    & 
    & 
    2.2\%
    &
    1.3\%
    \\
    $ \Delta^{\rm exact}_{\rm NNLO} $ (MSHT20)   
    & 
    &
    5.3\%
    &
    2.3\%
    \\
     $ \Delta^{\rm exact}_{\rm NNLO} $ (combination)   
    & 
    &
    3.3\%
    &
    2.3\%
    \\
    \midrule
    $ \Delta^{\rm app}_{\rm NNLO} $ (NNPDF4.0)   
    & 
    &
    0.2\%
    &
    0.2\%
    \\
    $ \Delta^{\rm app}_{\rm NNLO} $ (MSHT20)   
    & 
    &
    1.4\%
    &
    1.3\%
    \\
    $ \Delta^{\rm app}_{\rm NNLO} $ (combination)   
    & 
    &
    0.9\%
    &
    0.5\%
    \\
 \bottomrule
   \end{tabularx}
   \vspace{0.2cm}
   \caption{\small The total inclusive cross-section (in pb) for  Higgs  production in the gluon-fusion and VBF channels at the LHC Run 3 with $\sqrt{s}=13.6$~TeV.
   In all cases, the partonic matrix element is evaluated at N$^3$LO accuracy not including
   QED or electroweak corrections.
   The central scales are set to be $\mu_{F,R}=m_H/2$ (gluon fusion) and $\mu_{F,R}=Q_V$ (VBF).
The first uncertainty shown is the pure PDF uncertainty,
and the second  is the sum in quadrature of the PDF uncertainty and the MHOU on the hard cross-section evaluated with the 7-point prescription.
   The {\tt PDF4LHC21\_mc}~\cite{PDF4LHCWorkingGroup:2022cjn}  and {\tt  MSHT20xNNPDF40\_nnlo} PDF sets are at NNLO, while all the other PDF
   sets~\cite{McGowan:2022nag,NNPDF:2024nan} are determined at aN$^3$LO accuracy. 
   The error $\Delta^{\rm     exact}_{\rm NNLO}$, Eq.~(\ref{eq:PDFimpact_xsec_exact}),
   that is made when using NNLO PDFs with the aN$^3$LO matrix element, and
   its approximate estimate  $\Delta^{\rm app}_{\rm NNLO}$,
   Eq.~(\ref{eq:PDFimpact_xsec_approx}), are also provided both separately for NNPDF4.0, MSHT20, and for their combination in the case of the pure QCD fits.
   \label{table:pheno} }
\end{table}

\begin{table}[t!]
  \centering
  \footnotesize
   \renewcommand{\arraystretch}{2.0}
   \begin{tabularx}{\textwidth}
   {|X|l|l|l|l|}
     \toprule
  PDF set  &    
  pert. order (PDF)
  &  $\sigma(hW^+)$  
    &  $\sigma(hW^-)$  
      &  $\sigma(hZ)$
      \\
  \midrule
      {\tt PDF4LHC21\_mc}  
      &  
    NNLO$_{\rm QCD}$
      & 
       $0.944^{+1.6\%}_{-1.4\%}\,^{+1.6\%}_{-1.5\%}$
      &
      $0.593^{+1.5\%}_{-1.2\%}\,^{+1.5\%}_{-1.3\%}$
      & 
      $0.842^{+1.5\%}_{-1.0\%}\,^{+1.5\%}_{-1.1\%}$
      \\
         {\tt  MSHT20xNNPDF40\_nnlo}  
      &  
         NNLO$_{\rm QCD}$
      & 
       $0.957^{+1.9\%}_{-2.3\%}\,^{+1.9\%}_{-2.4\%}$
      &
      $0.601^{+1.5\%}_{-2.2\%}\,^{+1.6\%}_{-2.2\%}$
      & 
      $0.855^{+1.6\%}_{-2.0\%}\,^{+1.6\%}_{-2.0\%}$
      \\
         {\tt  MSHT20xNNPDF40\_nnlo\_qed}  
      &  
         NNLO$_{\rm QCD}\otimes {\rm NLO}_{\rm QED}$
      & 
       $0.952^{+2.1\%}_{-2.5\%}\,^{+2.1\%}_{-2.5\%}$
      &
      $0.598^{+1.8\%}_{-2.5\%}\,^{+1.8\%}_{-2.5\%}$
      & 
      $0.851^{+2.0\%}_{-2.4\%}\,^{+2.1\%}_{-2.4\%}$
      \\
      \midrule
        {\tt MSHT20xNNPDF40\_an3lo} 
      &  
       aN$^3$LO$_{\rm QCD}$
      &
       $0.961^{+1.4\%}_{-1.6\%}\,^{+1.4\%}_{-1.6\%}$
      &
       $0.604^{+1.2\%}_{-1.6\%}\,^{+1.2\%}_{-1.6\%}$
      &
       $0.859^{+1.4\%}_{-1.6\%}\,^{+1.4\%}_{-1.6\%}$
      \\
        {\tt MSHT20xNNPDF40\_an3lo\_qed} 
      &  
      aN$^3$LO$_{\rm QCD}\otimes {\rm NLO}_{\rm QED}$
      & 
      $0.955^{+1.7\%}_{-2.1\%}\,^{+1.7\%}_{-2.1\%}$
      &
      $0.601^{+1.4\%}_{-2.0\%}\,^{+1.4\%}_{-2.0\%}$
      & 
       $0.855^{+1.6\%}_{-2.0\%}\,^{+1.6\%}_{-2.0\%}$
      \\ \midrule
      {\tt NNPDF40\_an3lo\_as\_01180\_mhou}  &  
       aN$^3$LO$_{\rm QCD}$
      &
       $0.972^{+0.6\%}_{-0.6\%}\,^{+0.7\%}_{-0.7\%}$
      &
       $0.610^{+0.6\%}_{-0.6\%}\,^{+0.6\%}_{-0.7\%}$
      & 
       $0.869^{+0.5\%}_{-0.5\%}\,^{+0.6\%}_{-0.5\%}$
      \\
      {\tt NNPDF40\_an3lo\_as\_01180\_qed\_mhou}  & 
      aN$^3$LO$_{\rm QCD}\otimes {\rm NLO}_{\rm QED}$
      & 
       $0.969^{+0.6\%}_{-0.6\%}\,^{+0.6\%}_{-0.7\%}$
      &
      $0.608^{+0.5\%}_{-0.6\%}\,^{+0.6\%}_{-0.7\%}$
      & 
       $0.867^{+0.4\%}_{-0.4\%}\,^{+0.4\%}_{-0.5\%}$
      \\
       \midrule
      {\tt MSHT20an3lo\_as118} 
      & 
       aN$^3$LO$_{\rm QCD}$
      &
       $0.950^{+1.4\%}_{-1.4\%}\,^{+1.5\%}_{-1.5\%}$
      &
      $0.599^{+1.5\%}_{-1.5\%}\,^{+1.5\%}_{-1.5\%}$
      & 
       $0.849^{+1.4\%}_{-1.4\%}\,^{+1.4\%}_{-1.4\%}$
     \\
    {\tt MSHT20qed\_an3lo} 
    & 
    aN$^3$LO$_{\rm QCD}\otimes {\rm NLO}_{\rm QED}$
    &
      $0.941^{+1.5\%}_{-1.5\%}\,^{+1.5\%}_{-1.5\%}$
    &
    $0.595^{+1.6\%}_{-1.6\%}\,^{+1.6\%}_{-1.6\%}$
    & 
     $0.844^{+1.4\%}_{-1.4\%}\,^{+1.5\%}_{-1.5\%}$
     \\
      \midrule
       \midrule
    $ \Delta^{\rm exact}_{\rm NNLO} $ (NNPDF4.0)   
    & 
    & 
    0.5\%
    &
    0.3\%
    &
    0.3\%
    \\
    $ \Delta^{\rm exact}_{\rm NNLO} $ (MSHT20)   
    &
    &
    0.9\%
    &
    1.0\%
    &
    0.8\%
    \\
     $ \Delta^{\rm exact}_{\rm NNLO} $ (combination)   
    &
    &
    0.6\%
    &
    0.5\%
    &
    0.5\%
    \\
     \midrule
    $ \Delta^{\rm app}_{\rm NNLO} $ (NNPDF4.0)   
    & 
    &
    0.2\%
    &
    0.2\%
    &
    0.1\%
    \\
    $ \Delta^{\rm app}_{\rm NNLO} $ (MSHT20)   
    & 
    &
    0.8\%
    &
    0.9\%
    &
    1.1\%
    \\
    $ \Delta^{\rm app}_{\rm NNLO} $ (combination)   
    & 
    &
    0.1\%
    &
    0.2\%
    &
    0.3\%
    \\
 \bottomrule
   \end{tabularx}
   \vspace{0.2cm}
   \caption{\small Same as Table~\ref{table:pheno} for Higgs production in association with vector bosons. 
   \label{table:pheno2} }
\end{table}

Results are shown for  the  MSHT20 and
NNPDF4.0 aN$^3$LO sets both with and without QED corrections, and the
combined  MSHT20xNNPDF40 aN$^3$LO sets constructed here, also with and without QED
corrections. 
Results obtained by using perturbatively mismatched
NNLO PDFs together with the
N$^3$LO matrix element are also shown, both for the PDF4LHC21 and
 MSHT20xNNPDF40 NNLO combined sets. In order to visually assess the error
involved in this procedure, in Fig.~\ref{fig:higgs-ggF-n3lo-HXSWG} we
also show results  as a ratio of the result to that found using
mismatched NNLO PDFs, specifically from the  MSHT20xNNPDF40\_NNLO combined baseline set.

The  percentage error made when using mismatched
NNLO PDFs is
        \begin{equation}
     \label{eq:PDFimpact_xsec_exact}
      \Delta^{\rm exact}_{\rm NNLO} \equiv \Bigg| \frac{\sigma^{\rm N^3LO}_{\rm aN^3LO-PDF}
      - \sigma^{\rm N^3LO}_{\rm NNLO-PDF}}{\sigma^{\rm N^3LO}_{\rm aN^3LO-PDF}}\Bigg| \, .
        \end{equation}
 An approximate
          way of estimating this error before knowledge of aN$^3$LO
          PDFs, based on the behavior seen at one less perturbative
          order, was suggested in
          Refs.~\cite{Baglio:2022wzu,Anastasiou:2016cez} and it is
          given by
     \begin{equation}
      \label{eq:PDFimpact_xsec_approx}
\Delta^{\rm app}_{\rm NNLO} \equiv \frac{1}{2}\Bigg| \frac{\sigma^{\rm NNLO}_{\rm NNLO-PDF}
      - \sigma^{\rm NNLO}_{\rm NLO-PDF}}{\sigma^{\rm NNLO}_{\rm
          NNLO-PDF}}\Bigg| \, .
     \end{equation}
 The values of $\Delta^{\rm exact}_{\rm NNLO}$ and $\Delta^{\rm
       app}_{\rm NNLO}$ obtained for each process are also included in
     Tables~\ref{table:pheno}-\ref{table:pheno2}, both for NNPDF4.0,
     MSHT20, and the combined PDF sets constructed here, using in each case
     PDFs at the required perturbative orders from the same set. For the computation of
     $\Delta^{\rm app}_{\rm NNLO}$ for the combined set we use as a value for  $\sigma^{\rm
       NNLO}_{\rm NLO-PDF}$ the average of the cross-sections computed using NNPDF4.0
     and MSHT20 NLO PDFs.

It is clear that, as already noticed from the comparison of parton
luminosities of
Figs.~\ref{fig:lumi-n3lo-PDF4HXSWG}-\ref{fig:lumi-nnlo-PDF4HXSWG}, the
effect of using aN$^3$LO PDFs is most substantial for gluon fusion and vector
boson fusion, which depend on the gluon-gluon and quark-quark
luminosity respectively, and is small for associated $hV$ production, which depends on the
quark-antiquark luminosity. 
It follows that for gluon fusion and VBF using mismatched NNLO
PDFs leads to a large error $\Delta^{\rm exact}_{\rm NNLO}$.
Furthermore, the approximate estimate of this error  $\Delta^{\rm
  app}_{\rm NNLO}$ is very unreliable, and specifically for gluon
fusion and VBF it underestimates the true $\Delta^{\rm exact}_{\rm NNLO}$ 
significantly.
The fact  that for gluon fusion $\Delta^{\rm
  app}_{\rm NNLO}$ is rather smaller than $\Delta^{\rm exact}_{\rm
  NNLO}$ means that the difference between using N$^3$LO and NNLO PDFs
is significantly larger than one may expect comparing the change at
the previous order, thus belying
expectations~\cite{Forte:2013mda,Anastasiou:2016cez} that the impact of
N$^3$LO PDFs would be very small based on the behavior of the
previous orders. This is a consequence of the behavior of the gluon
luminosity discussed at the end of Sect.~\ref{sec:comb}.

Indeed,  for gluon fusion
  there is a compensation between the hard
  cross-section and the gluon PDF when both are used at N$^3$LO, with
  the former leading to a significant increase which is then largely
  eliminated by the reduced gluon distribution in the correctly
  matched PDFs.
  For VBF, instead, the use of aN$^3$LO PDFs leads
  to a clear increase of $2.5\%$ compared to using NNLO PDFs in
  all cases, much larger than the fixed-order uncertainty.

The choice of a specific PDF (either of the individual sets or any of
the combinations) entering the calculation has a relatively small effect on
VBF, either at NNLO or aN$^3$LO. For gluon fusion this choice has a
negligible effect at NNLO, while at aN$^3$LO, as repeatedly observed,
a stronger suppression is found using MSHT20 than NNPDF4.0 PDFs, with
the combination providing a value in between and an uncertainty
encompassing the different degrees of suppression. Hence in both
cases, the error made using NNLO PDFs is larger than the PDF
uncertainties, and also larger than the spread of aN$^3$LO results. 
The picture is somewhat different for associate Higgs production.
For this process, the impact of using aN$^3$LO vs. NNLO PDFs is negligible, so the
dominant difference comes from the choice of PDF set, though results
found using any of the PDF sets or combinations considered here all
agree within uncertainties.
  
Also, we observe that   
for Higgs production in gluon fusion the effect of QED corrections, again as already seen in
luminosities, is also not negligible, though smaller than that of
using properly matched PDFs.

It is also interesting to compare results obtained using the MC combination of aN$^3$LO PDF sets to those that would be found by performing the textbook weighted average and uncertainty.
For gluon fusion the weighted combination (with weights  $1/\delta_{\rm pdf}^2$) gives
 \begin{align}\label{eq:comb}
\sigma_{\rm ggF}^{\rm comb}&=  
45.30\pm 0.28_{\rm pdf}~\hbox{pb} \quad \hbox{(pure QCD)}   
\\
\label{eq:comb_qed}
  \sigma_{\rm ggF}^{\rm comb}      &=   44.43\pm 0.26_{\rm pdf}~\hbox{pb} \quad \hbox{(QCD+QED)}.
\end{align}
This is to be compared with the result found using the  MSHT20xNNPDF40 aN$^3$LO sets (both pure QCD and QCD$\otimes$QED) from Table~\ref{table:pheno}, namely
  \begin{align}\label{eq:comb1}
\sigma_{\rm ggF}^{\rm MSHTxNNPDF}&= 
44.86\pm 0.90_{\rm pdf}~\hbox{pb} 
\quad \hbox{(pure QCD)}   
\\\label{eq:comb_qed1}
  \sigma_{\rm ggF}^{\rm MSHTxNNPDF\_qed}  
  &=   44.20\pm 0.75_{\rm pdf}~\hbox{pb} \quad \hbox{(QCD+QED)}.
\end{align}
where in both cases we consider only the PDF uncertainty.
It is clear that the latter result is significantly more conservative,
in that it treats the two existing sets on the same footing and leads
to a correspondingly rather larger uncertainty.  
Eqs.~(\ref{eq:comb1})--(\ref{eq:comb_qed1}) should be considered our
best result for Higgs production in gluon fusion, although we emphasize that the difference between this and the weighted average remains lower than the difference between the predictions based on NNLO and aN$^3$LO PDFs.

Finally, as mentioned in the introduction, we have assessed the effect of using the preliminary variants of the aN$^3$LO PDF sets presented here, and recently constructed by MSHT~\cite{mshttalk,mshttalk2} and NNPDF~\cite{nnpdftalk} using the parametrization of aN$^3$LO splitting functions of Ref.~\cite{Falcioni:2024qpd}, that includes all the perturbative information that is currently available, and also using for both sets the same N$^3$LO heavy quark transition matrix element as used by NNPDF4.0. These preliminary variants have been also used in order to produce a variant combined set.
Very moderate differences are found: specifically,  for NNPDF,
the central value of the gluon-fusion cross-section in the aN$^3$LO baseline fit of 45.49 pb (see Table~\ref{table:pheno}) becomes 
45.25 pb, namely a shift
of $-0.5\%$, while for MSHT the aN$^3$LO baseline result of 44.08 pb becomes 44.26 pb, a shift of $+0.4\%$ (when only updated the splitting functions the respective value is 44.38 pb, i.e. a shift of $0.7\%$). In the case of the VBF cross-sections, we have verified that the impact of the updated splitting functions is negligible, i.e. of order $0.1\%$. 
Therefore, the difference between the central predictions of the aN$^3$LO fits of NNPDF and MSHT is reduced by around 
$1\%$ when the
same parametrization of splitting functions from~\cite{Falcioni:2024qpd} is used. However, the average value obtained from a combined set of PDFs is almost unchanged. Moreover, because the updated PDFs get closer, the uncertainty found with this updated combination is reduced, consistent with the reduction in theoretical uncertainty due to better knowledge of  N$^3$LO splitting functions and transition matrix elements. In summary, the updated combination leads to an almost identical central prediction and a somewhat reduced uncertainty. 

Hence,   we conclude that the results using the combination presented here, based on published PDF sets, can currently be relied upon to produce a reliable set of predictions and a conservative uncertainty estimate. Similarly, the individual published aN$^3$LO PDF sets can be taken to be reliable, as updated information results in very similar predictions well within the uncertainty, which again can be taken as conservative as improved N$^3$LO knowledge would reduce theoretical uncertainty. On the other hand, we conclude that adopting a common set of N$^3$LO splitting function does improve the overall consistency between the MSHT20 and NNPDF4.0 aN$^3$LO sets, especially for the gluon-gluon luminosity. As more information on N$^3$LO terms becomes available, it will be incorporated in future PDF sets, and  PDFs based on common settings could be considered for future combinations. For the time being the existing sets may be reliably used, until they will be replaced by some newer version including updates based on improvements in N$^3$LO knowledge but also potentially on increased data and other methodological changes.

\section{Conclusion}
\label{sec:conclusion}

In this note we have shown that the use of aN$^3$LO PDFs is mandatory for accurate N$^3$LO phenomenology at the LHC, 
focusing on the case of Higgs production cross-sections.
For Higgs production in gluon fusion and in vector boson fusion  the
impact of switching from NNLO to aN$^3$LO PDFs is rather  larger than
the PDF uncertainty, and also larger (for VBF much larger) than the
difference between different PDF sets. Indeed,
we have shown that differences between existing aN$^3$LO PDF sets are moderate, and generally smaller than the difference between NNLO and aN$^3$LO. We have verified that updates due to improved knowledge of N$^3$LO splitting functions and transition matrix elements does not alter this conclusion. 

Our results illustrate the limitations of N$^3$LO phenomenology at the LHC using NNLO PDFs.
In order to deal with these limitations, we have constructed  combined
aN$^3$LO PDF sets, both pure QCD and
QCD+QED, which allow for the computation of
phenomenological predictions with a more conservative estimate of the
uncertainty, as is appropriate when combining predictions that are not fully compatible. 
Clearly,  a full N$^3$LO PDF determination will only be possible once all processes currently used for PDF determination are known at this order, which is likely to take some time. In particular, the calculation of the massive corrections to the DIS coefficient functions and the full set of N$^3$LO corrections to hadronic processes will almost certainly take many years to compute. 
In the meantime, our results clearly demonstrate that the available aN$^3$LO PDFs represent the most accurate option for the deployment of N$^3$LO calculations for LHC physics.

\vspace{1cm}
The {\tt  MSHT20xNNPDF40\_an3lo} and {\tt  MSHT20xNNPDF40\_an3lo\_qed}
PDF sets presented in this note have been submitted to  the {\sc
  LHAPDF} repository~\cite{Buckley:2014ana}. They are also made
available (alongside the corresponding NNLO baseline sets) at the link: 
\begin{center}
    \href{https://zenodo.org/records/13843626}{https://zenodo.org/records/13843626}.
\end{center}
\bigskip

\subsection*{Acknowledgments} 

We thank the convenors of Working Group 1 of the LHC Higgs Cross-Section Working Group (LHCHXSWG), in particular the gluon-fusion subgroup, for constructive feedback on the first draft of this document.
We are also grateful to Amanda Cooper-Sarkar for questions and critical input, and members of the CT collaboration, in particular Joey Huston and Pavel Nadolsky, for discussion concerning N$^3$LO PDFs and their uncertainties. 

S.~F. performed this work in part at the Aspen Center for Physics, which is supported by National
Science Foundation Grant PHY-2210452, and is partially supported by an Italian PRIN2022 grant. R.~D.~B, L.~D.~D. and R.~S. thank the Science and Technology Facilities Council (STFC) for support via grant awards ST/T000600/1 and ST/X000494/1. L.~H.-L. and R.~S.~T. thank the Science and Technology Facilities Council (STFC) for support via grant awards ST/T000856/1 and ST/X000516/1. T.~C. acknowledges that this project has received funding from the European Research Council (ERC) under the European Union’s Horizon 2020 research and innovation programme (Grant agreement No. 101002090 COLORFREE). E.~R.~N. is supported by the Italian Ministry of University and Research (MUR) through the “Rita Levi-Montalcini” Program.
F.~H. is supported by the Academy of Finland project 358090 and is funded as a part of the Center of Excellence in Quark Matter of the Academy of Finland, project 346326. 
 M.U. is supported by the European Research Council under the European
Union’s Horizon 2020 research and innovation Programme (grant agreement n.950246), and partially by the STFC consolidated grant ST/T000694/1 and ST/X000664/1.

\providecommand{\href}[2]{#2}\begingroup\raggedright\endgroup


\begin{thebibliography}{10}

\bibitem{Bailey:2020ooq}
S.~Bailey, T.~Cridge, L.~A. Harland-Lang, A.~D. Martin, and R.~S. Thorne, {\it
  {Parton distributions from LHC, HERA, Tevatron and fixed target data: MSHT20
  PDFs}},  {\em Eur. Phys. J. C} {\bf 81} (2021), no.~4 341,
  [\href{http://arxiv.org/abs/2012.04684}{{\tt arXiv:2012.04684}}].

\bibitem{NNPDF:2021njg}
{\bf NNPDF} Collaboration, R.~D. Ball et~al., {\it {The path to proton
  structure at 1\% accuracy}},  {\em Eur. Phys. J. C} {\bf 82} (2022), no.~5
  428, [\href{http://arxiv.org/abs/2109.02653}{{\tt arXiv:2109.02653}}].

\bibitem{McGowan:2022nag}
J.~McGowan, T.~Cridge, L.~A. Harland-Lang, and R.~S. Thorne, {\it {Approximate
  N$^{3}$LO parton distribution functions with theoretical uncertainties:
  MSHT20aN$^3$LO PDFs}},  {\em Eur. Phys. J. C} {\bf 83} (2023), no.~3 185,
  [\href{http://arxiv.org/abs/2207.04739}{{\tt arXiv:2207.04739}}]. [Erratum:
  Eur.Phys.J.C 83, 302 (2023)].

\bibitem{NNPDF:2024nan}
{\bf NNPDF} Collaboration, R.~D. Ball et~al., {\it {The path to $\hbox
  {N}^3\hbox {LO}$ parton distributions}},  {\em Eur. Phys. J. C} {\bf 84}
  (2024), no.~7 659, [\href{http://arxiv.org/abs/2402.18635}{{\tt
  arXiv:2402.18635}}].

\bibitem{Cridge:2023ryv}
T.~Cridge, L.~A. Harland-Lang, and R.~S. Thorne, {\it {Combining QED and
  approximate ${\rm N}^3$LO QCD corrections in a global PDF fit:
  MSHT20qed\_an3lo PDFs}},  {\em SciPost Phys.} {\bf 17} (2024), no.~1 026,
  [\href{http://arxiv.org/abs/2312.07665}{{\tt arXiv:2312.07665}}].

\bibitem{Barontini:2024dyb}
A.~Barontini, N.~Laurenti, and J.~Rojo, {\it {NNPDF4.0 aN$^3$LO PDFs with QED
  corrections}},  in {\em {31st International Workshop on Deep-Inelastic
  Scattering and Related Subjects}}, 6, 2024.
\newblock \href{http://arxiv.org/abs/2406.01779}{{\tt arXiv:2406.01779}}.

\bibitem{Moch:2017uml}
S.~Moch, B.~Ruijl, T.~Ueda, J.~A.~M. Vermaseren, and A.~Vogt, {\it {Four-Loop
  Non-Singlet Splitting Functions in the Planar Limit and Beyond}},  {\em JHEP}
  {\bf 10} (2017) 041, [\href{http://arxiv.org/abs/1707.08315}{{\tt
  arXiv:1707.08315}}].

\bibitem{Davies:2016jie}
J.~Davies, A.~Vogt, B.~Ruijl, T.~Ueda, and J.~A.~M. Vermaseren, {\it {Large-nf
  contributions to the four-loop splitting functions in QCD}},  {\em Nucl.
  Phys. B} {\bf 915} (2017) 335--362,
  [\href{http://arxiv.org/abs/1610.07477}{{\tt arXiv:1610.07477}}].

\bibitem{Gehrmann:2023cqm}
T.~Gehrmann, A.~von Manteuffel, V.~Sotnikov, and T.-Z. Yang, {\it {Complete $
  {N}_f^2 $ contributions to four-loop pure-singlet splitting functions}},
  {\em JHEP} {\bf 01} (2024) 029, [\href{http://arxiv.org/abs/2308.07958}{{\tt
  arXiv:2308.07958}}].

\bibitem{Gehrmann:2023iah}
T.~Gehrmann, A.~von Manteuffel, V.~Sotnikov, and T.-Z. Yang, {\it {The $N_f
  C_F^3$ contribution to the non-singlet splitting function at four-loop
  order}},  {\em Phys. Lett. B} {\bf 849} (2024) 138427,
  [\href{http://arxiv.org/abs/2310.12240}{{\tt arXiv:2310.12240}}].

\bibitem{Falcioni:2023tzp}
G.~Falcioni, F.~Herzog, S.~Moch, J.~Vermaseren, and A.~Vogt, {\it {The double
  fermionic contribution to the four-loop quark-to-gluon splitting function}},
  {\em Phys. Lett. B} {\bf 848} (2024) 138351,
  [\href{http://arxiv.org/abs/2310.01245}{{\tt arXiv:2310.01245}}].

\bibitem{Moch:2018wjh}
S.~Moch, B.~Ruijl, T.~Ueda, J.~A.~M. Vermaseren, and A.~Vogt, {\it {On quartic
  colour factors in splitting functions and the gluon cusp anomalous
  dimension}},  {\em Phys. Lett. B} {\bf 782} (2018) 627--632,
  [\href{http://arxiv.org/abs/1805.09638}{{\tt arXiv:1805.09638}}].

\bibitem{Moch:2021qrk}
S.~Moch, B.~Ruijl, T.~Ueda, J.~A.~M. Vermaseren, and A.~Vogt, {\it {Low moments
  of the four-loop splitting functions in QCD}},  {\em Phys. Lett. B} {\bf 825}
  (2022) 136853, [\href{http://arxiv.org/abs/2111.15561}{{\tt
  arXiv:2111.15561}}].

\bibitem{Falcioni:2023luc}
G.~Falcioni, F.~Herzog, S.~Moch, and A.~Vogt, {\it {Four-loop splitting
  functions in QCD \textendash{} The quark-quark case}},  {\em Phys. Lett. B}
  {\bf 842} (2023) 137944, [\href{http://arxiv.org/abs/2302.07593}{{\tt
  arXiv:2302.07593}}].

\bibitem{Falcioni:2023vqq}
G.~Falcioni, F.~Herzog, S.~Moch, and A.~Vogt, {\it {Four-loop splitting
  functions in QCD \textendash{} The gluon-to-quark case}},  {\em Phys. Lett.
  B} {\bf 846} (2023) 138215, [\href{http://arxiv.org/abs/2307.04158}{{\tt
  arXiv:2307.04158}}].

\bibitem{Moch:2023tdj}
S.~Moch, B.~Ruijl, T.~Ueda, J.~Vermaseren, and A.~Vogt, {\it {Additional
  moments and x-space approximations of four-loop splitting functions in QCD}},
   {\em Phys. Lett. B} {\bf 849} (2024) 138468,
  [\href{http://arxiv.org/abs/2310.05744}{{\tt arXiv:2310.05744}}].

\bibitem{Falcioni:2024xyt}
G.~Falcioni, F.~Herzog, S.~Moch, A.~Pelloni, and A.~Vogt, {\it {Four-loop
  splitting functions in QCD \textendash{} The quark-to-gluon case}},  {\em
  Phys. Lett. B} {\bf 856} (2024) 138906,
  [\href{http://arxiv.org/abs/2404.09701}{{\tt arXiv:2404.09701}}].

\bibitem{Falcioni:2024xav}
G.~Falcioni, F.~Herzog, S.~Moch, and S.~Van~Thurenhout, {\it {Constraints for
  twist-two alien operators in QCD}},
  \href{http://arxiv.org/abs/2409.02870}{{\tt arXiv:2409.02870}}.

\bibitem{Falcioni:2024qpd}
G.~Falcioni, F.~Herzog, S.~Moch, A.~Pelloni, and A.~Vogt, {\it {Four-loop
  splitting functions in QCD -- The gluon-gluon case --}},
  \href{http://arxiv.org/abs/2410.08089}{{\tt arXiv:2410.08089}}.

\bibitem{Kawamura:2012cr}
H.~Kawamura, N.~A. Lo~Presti, S.~Moch, and A.~Vogt, {\it {On the
  next-to-next-to-leading order QCD corrections to heavy-quark production in
  deep-inelastic scattering}},  {\em Nucl. Phys. B} {\bf 864} (2012) 399--468,
  [\href{http://arxiv.org/abs/1205.5727}{{\tt arXiv:1205.5727}}].

\bibitem{Bierenbaum:2009mv}
I.~Bierenbaum, J.~Bl\"umlein, and S.~Klein, {\it {Mellin Moments of the
  $O(\alpha_s^3)$ Heavy Flavor Contributions to unpolarized Deep-Inelastic
  Scattering at $Q^2 \gg m^2$ and Anomalous Dimensions}},  {\em Nucl. Phys. B}
  {\bf 820} (2009) 417--482, [\href{http://arxiv.org/abs/0904.3563}{{\tt
  arXiv:0904.3563}}].

\bibitem{Ablinger:2014vwa}
J.~Ablinger, A.~Behring, J.~Bl\"umlein, A.~De~Freitas, A.~Hasselhuhn, A.~von
  Manteuffel, M.~Round, C.~Schneider, and F.~Wi\ss{}brock, {\it {The 3-Loop
  Non-Singlet Heavy Flavor Contributions and Anomalous Dimensions for the
  Structure Function $F_2(x,Q^2)$ and Transversity}},  {\em Nucl. Phys. B} {\bf
  886} (2014) 733--823, [\href{http://arxiv.org/abs/1406.4654}{{\tt
  arXiv:1406.4654}}].

\bibitem{Ablinger:2014nga}
J.~Ablinger, A.~Behring, J.~Bl\"umlein, A.~De~Freitas, A.~von Manteuffel, and
  C.~Schneider, {\it {The 3-loop pure singlet heavy flavor contributions to the
  structure function $F_2(x,Q^2)$ and the anomalous dimension}},  {\em Nucl.
  Phys. B} {\bf 890} (2014) 48--151,
  [\href{http://arxiv.org/abs/1409.1135}{{\tt arXiv:1409.1135}}].

\bibitem{Blumlein:2021enk}
J.~Bl\"umlein, P.~Marquard, C.~Schneider, and K.~Sch\"onwald, {\it {The
  three-loop unpolarized and polarized non-singlet anomalous dimensions from
  off shell operator matrix elements}},  {\em Nucl. Phys. B} {\bf 971} (2021)
  115542, [\href{http://arxiv.org/abs/2107.06267}{{\tt arXiv:2107.06267}}].

\bibitem{Ablinger:2014uka}
J.~Ablinger, J.~Bl\"umlein, A.~De~Freitas, A.~Hasselhuhn, A.~von Manteuffel,
  M.~Round, and C.~Schneider, {\it {The $O(\alpha_s^3 T_F^2)$ Contributions to
  the Gluonic Operator Matrix Element}},  {\em Nucl. Phys. B} {\bf 885} (2014)
  280--317, [\href{http://arxiv.org/abs/1405.4259}{{\tt arXiv:1405.4259}}].

\bibitem{Ablinger:2014tla}
J.~Ablinger, J.~Bl\"umlein, A.~De~Freitas, A.~Hasselhuhn, A.~von Manteuffel,
  M.~Round, and C.~Schneider, {\it {3-loop Massive $O(T_F^2)$ Contributions to
  the DIS Operator Matrix Element $A_{gg}$}},  {\em Nucl. Part. Phys. Proc.}
  {\bf 258-259} (2015) 37--40, [\href{http://arxiv.org/abs/1409.1435}{{\tt
  arXiv:1409.1435}}].

\bibitem{ablinger:agq}
J.~{Ablinger}, J.~{Bl{\"u}mlein}, A.~{De Freitas}, A.~{Hasselhuhn}, A.~{von
  Manteuffel}, M.~{Round}, C.~{Schneider}, and F.~{Wi{\ss}brock}, {\it {The
  transition matrix element A$_{gq}$(N) of the variable flavor number scheme at
  O({\ensuremath{\alpha}}s3)}},  {\em Nuclear Physics B} {\bf 882} (May, 2014)
  263--288, [\href{http://arxiv.org/abs/1402.0359}{{\tt arXiv:1402.0359}}].

\bibitem{Ablinger:2022wbb}
J.~Ablinger, A.~Behring, J.~Bl\"umlein, A.~De~Freitas, A.~Goedicke, A.~von
  Manteuffel, C.~Schneider, and K.~Sch\"onwald, {\it {The unpolarized and
  polarized single-mass three-loop heavy flavor operator matrix elements
  A$_{gg,Q}$ and \ensuremath{\Delta}A$_{gg,Q}$}},  {\em JHEP} {\bf 12} (2022)
  134, [\href{http://arxiv.org/abs/2211.05462}{{\tt arXiv:2211.05462}}].

\bibitem{Ablinger:2023ahe}
J.~Ablinger, A.~Behring, J.~Bl\"umlein, A.~De~Freitas, A.~von Manteuffel,
  C.~Schneider, and K.~Sch\"onwald, {\it {The first\textendash{}order
  factorizable contributions to the three\textendash{}loop massive operator
  matrix elements AQg(3) and \ensuremath{\Delta}AQg(3)}},  {\em Nucl. Phys. B}
  {\bf 999} (2024) 116427, [\href{http://arxiv.org/abs/2311.00644}{{\tt
  arXiv:2311.00644}}].

\bibitem{Ablinger:2024xtt}
J.~Ablinger, A.~Behring, J.~Bl\"umlein, A.~De~Freitas, A.~von Manteuffel,
  C.~Schneider, and K.~Sch\"onwald, {\it {The non-first-order-factorizable
  contributions to the three-loop single-mass operator matrix elements
  $A_{Qg}^{(3)}$ and $\Delta A_{Qg}^{(3)}$}},
  \href{http://arxiv.org/abs/2403.00513}{{\tt arXiv:2403.00513}}.

\bibitem{Vermaseren:2005qc}
J.~A.~M. Vermaseren, A.~Vogt, and S.~Moch, {\it {The third-order QCD
  corrections to deep-inelastic scattering by photon exchange}},  {\em Nucl.
  Phys.} {\bf B724} (2005) 3, [\href{http://arxiv.org/abs/hep-ph/0504242}{{\tt
  hep-ph/0504242}}].

\bibitem{Blumlein:2022gpp}
J.~Bl\"umlein, P.~Marquard, C.~Schneider, and K.~Sch\"onwald, {\it {The
  massless three-loop Wilson coefficients for the deep-inelastic structure
  functions F$_{2}$, F$_{L}$, xF$_{3}$ and g$_{1}$}},  {\em JHEP} {\bf 11}
  (2022) 156, [\href{http://arxiv.org/abs/2208.14325}{{\tt arXiv:2208.14325}}].

\bibitem{Caola:2022ayt}
F.~Caola, W.~Chen, C.~Duhr, X.~Liu, B.~Mistlberger, F.~Petriello, G.~Vita, and
  S.~Weinzierl, {\it {The Path forward to N$^3$LO}},  in {\em {Snowmass 2021}},
  3, 2022.
\newblock \href{http://arxiv.org/abs/2203.06730}{{\tt arXiv:2203.06730}}.

\bibitem{Duhr:2020sdp}
C.~Duhr, F.~Dulat, and B.~Mistlberger, {\it {Charged current Drell-Yan
  production at N$^{3}$LO}},  {\em JHEP} {\bf 11} (2020) 143,
  [\href{http://arxiv.org/abs/2007.13313}{{\tt arXiv:2007.13313}}].

\bibitem{Duhr:2020seh}
C.~Duhr, F.~Dulat, and B.~Mistlberger, {\it {Drell-Yan Cross Section to Third
  Order in the Strong Coupling Constant}},  {\em Phys. Rev. Lett.} {\bf 125}
  (2020), no.~17 172001, [\href{http://arxiv.org/abs/2001.07717}{{\tt
  arXiv:2001.07717}}].

\bibitem{duhr:DY2021}
C.~Duhr and B.~Mistlberger, {\it {Lepton-pair production at hadron colliders at
  N$^{3}$LO in QCD}},  {\em JHEP} {\bf 03} (2022) 116,
  [\href{http://arxiv.org/abs/2111.10379}{{\tt arXiv:2111.10379}}].

\bibitem{Gehrmann:DYN3LO}
X.~Chen, T.~Gehrmann, N.~Glover, A.~Huss, T.-Z. Yang, and H.~X. Zhu, {\it
  {Dilepton Rapidity Distribution in Drell-Yan Production to Third Order in
  QCD}},  {\em Phys. Rev. Lett.} {\bf 128} (2022), no.~5 052001,
  [\href{http://arxiv.org/abs/2107.09085}{{\tt arXiv:2107.09085}}].

\bibitem{Ball:2012wy}
R.~D. Ball, S.~Carrazza, L.~Del~Debbio, S.~Forte, J.~Gao, et~al., {\it {Parton
  Distribution Benchmarking with LHC Data}},  {\em JHEP} {\bf 1304} (2013) 125,
  [\href{http://arxiv.org/abs/1211.5142}{{\tt arXiv:1211.5142}}].

\bibitem{Ball:2021icz}
R.~D. Ball and R.~L. Pearson, {\it {Correlation of theoretical uncertainties in
  PDF fits and theoretical uncertainties in predictions}},  {\em Eur. Phys. J.
  C} {\bf 81} (2021), no.~9 830, [\href{http://arxiv.org/abs/2105.05114}{{\tt
  arXiv:2105.05114}}].

\bibitem{Davies:2022ofz}
J.~Davies, C.~H. Kom, S.~Moch, and A.~Vogt, {\it {Resummation of small-x double
  logarithms in QCD: inclusive deep-inelastic scattering}},  {\em JHEP} {\bf
  08} (2022) 135, [\href{http://arxiv.org/abs/2202.10362}{{\tt
  arXiv:2202.10362}}].

\bibitem{Andersen:2024czj}
J.~Andersen et~al., {\it {Les Houches 2023: Physics at TeV Colliders: Standard
  Model Working Group Report}},  in {\em {Physics of the TeV Scale and Beyond
  the Standard Model}: {Intensifying the Quest for New Physics}}, 6, 2024.
\newblock \href{http://arxiv.org/abs/2406.00708}{{\tt arXiv:2406.00708}}.

\bibitem{Cooper-Sarkar:2024crx}
A.~Cooper-Sarkar, T.~Cridge, F.~Giuli, L.~A. Harland-Lang, F.~Hekhorn,
  J.~Huston, G.~Magni, S.~Moch, and R.~S. Thorne, {\it {A Benchmarking of QCD
  Evolution at Approximate $N^3LO$}},
  \href{http://arxiv.org/abs/2406.16188}{{\tt arXiv:2406.16188}}.

\bibitem{Thorne:2024npj}
R.~S. Thorne, T.~Cridge, and L.~Harland-Lang, {\it {MSHT Updates 2024}},  in
  {\em {31st International Workshop on Deep-Inelastic Scattering and Related
  Subjects}}, 8, 2024.
\newblock \href{http://arxiv.org/abs/2408.10008}{{\tt arXiv:2408.10008}}.

\bibitem{mshttalk}
R.~Thorne, {\it
  \url{https://indico.cern.ch/event/1435677/contributions/6133441/attachments/2977567/5242015/PDF4LHC2024.pdf}},
  . ``MSHT PDF Update'', PDF4LHC Workshop 2024.

\bibitem{mshttalk2}
T.~Cridge, {\it
  \url{https://indico.cern.ch/event/1436959/contributions/6352248/author/9387795}},
  . "Combination of aN3LO PDFs and implications for Higgs production
  cross-sections at the LHC", DIS2025.

\bibitem{nnpdftalk}
G.~Magni, {\it
  \url{https://indico.cern.ch/event/1435677/contributions/6246030/attachments/2977793/5242448/PDF4LHC_an3lo_02_12_24.pdf}},
  . ``Combination of aN3LO PDFs and implications for Higgs production'',
  PDF4LHC Workshop 2024.

\bibitem{Baglio:2022wzu}
J.~Baglio, C.~Duhr, B.~Mistlberger, and R.~Szafron, {\it {Inclusive production
  cross sections at N$^{3}$LO}},  {\em JHEP} {\bf 12} (2022) 066,
  [\href{http://arxiv.org/abs/2209.06138}{{\tt arXiv:2209.06138}}].

\bibitem{Anastasiou:2016cez}
C.~Anastasiou, C.~Duhr, F.~Dulat, E.~Furlan, T.~Gehrmann, F.~Herzog,
  A.~Lazopoulos, and B.~Mistlberger, {\it {High precision determination of the
  gluon fusion Higgs boson cross-section at the LHC}},  {\em JHEP} {\bf 05}
  (2016) 058, [\href{http://arxiv.org/abs/1602.00695}{{\tt arXiv:1602.00695}}].

\bibitem{Butterworth:2015oua}
J.~Butterworth et~al., {\it {PDF4LHC recommendations for LHC Run II}},  {\em J.
  Phys.} {\bf G43} (2016) 023001, [\href{http://arxiv.org/abs/1510.03865}{{\tt
  arXiv:1510.03865}}].

\bibitem{NNPDF:2024dpb}
{\bf NNPDF} Collaboration, R.~D. Ball et~al., {\it {Determination of the theory
  uncertainties from missing higher orders on NNLO parton distributions with
  percent accuracy}},  {\em Eur. Phys. J. C} {\bf 84} (2024), no.~5 517,
  [\href{http://arxiv.org/abs/2401.10319}{{\tt arXiv:2401.10319}}].

\bibitem{PDF4LHCWorkingGroup:2022cjn}
{\bf PDF4LHC Working Group} Collaboration, R.~D. Ball et~al., {\it {The
  PDF4LHC21 combination of global PDF fits for the LHC Run III}},  {\em J.
  Phys. G} {\bf 49} (2022), no.~8 080501,
  [\href{http://arxiv.org/abs/2203.05506}{{\tt arXiv:2203.05506}}].

\bibitem{Cridge:2021qjj}
{\bf PDF4LHC21 combination group} Collaboration, T.~Cridge, {\it {PDF4LHC21:
  Update on the benchmarking of the CT, MSHT and NNPDF global PDF fits}},  {\em
  SciPost Phys. Proc.} {\bf 8} (2022) 101,
  [\href{http://arxiv.org/abs/2108.09099}{{\tt arXiv:2108.09099}}].

\bibitem{Erler:2020bif}
J.~Erler and R.~Ferro-Hern\'andez, {\it {Alternative to the application of PDG
  scale factors}},  {\em Eur. Phys. J. C} {\bf 80} (2020), no.~6 541,
  [\href{http://arxiv.org/abs/2004.01219}{{\tt arXiv:2004.01219}}].

\bibitem{LHCHiggsCrossSectionWorkingGroup:2011wcg}
{\bf LHC Higgs Cross Section Working Group} Collaboration, S.~Dittmaier et~al.,
  {\it {Handbook of LHC Higgs Cross Sections: 1. Inclusive Observables}},
  \href{http://arxiv.org/abs/1101.0593}{{\tt arXiv:1101.0593}}.

\bibitem{Carrazza:2015hva}
S.~Carrazza, J.~I. Latorre, J.~Rojo, and G.~Watt, {\it {A compression algorithm
  for the combination of PDF sets}},  {\em Eur. Phys. J.} {\bf C75} (2015) 474,
  [\href{http://arxiv.org/abs/1504.06469}{{\tt arXiv:1504.06469}}].

\bibitem{Watt:2012tq}
G.~Watt and R.~S. Thorne, {\it {Study of Monte Carlo approach to experimental
  uncertainty propagation with MSTW 2008 PDFs}},  {\em JHEP} {\bf 1208} (2012)
  052, [\href{http://arxiv.org/abs/1205.4024}{{\tt arXiv:1205.4024}}].

\bibitem{Ball:2021dab}
R.~D. Ball, S.~Forte, and R.~Stegeman, {\it {Correlation and combination of
  sets of parton distributions}},  {\em Eur. Phys. J. C} {\bf 81} (2021),
  no.~11 1046, [\href{http://arxiv.org/abs/2110.08274}{{\tt
  arXiv:2110.08274}}].

\bibitem{NNPDF:2024djq}
{\bf NNPDF} Collaboration, R.~D. Ball et~al., {\it {Photons in the proton:
  implications for the LHC}},  \href{http://arxiv.org/abs/2401.08749}{{\tt
  arXiv:2401.08749}}.

\bibitem{Bonvini:2014jma}
M.~Bonvini, R.~D. Ball, S.~Forte, S.~Marzani, and G.~Ridolfi, {\it {Updated
  Higgs cross section at approximate N$^3$LO}},  {\em J. Phys.} {\bf G41}
  (2014) 095002, [\href{http://arxiv.org/abs/1404.3204}{{\tt
  arXiv:1404.3204}}].

\bibitem{Bonvini:2013kba}
M.~Bonvini, {\it {An approximate N$^3$LO cross section for Higgs production in
  gluon fusion}},  {\em EPJ Web Conf.} {\bf 60} (2013) 12008,
  [\href{http://arxiv.org/abs/1306.6633}{{\tt arXiv:1306.6633}}].

\bibitem{Dreyer:2016oyx}
F.~A. Dreyer and A.~Karlberg, {\it {Vector-Boson Fusion Higgs Production at
  Three Loops in QCD}},  {\em Phys. Rev. Lett.} {\bf 117} (2016), no.~7 072001,
  [\href{http://arxiv.org/abs/1606.00840}{{\tt arXiv:1606.00840}}].

\bibitem{Dreyer:2018qbw}
F.~A. Dreyer and A.~Karlberg, {\it {Vector-Boson Fusion Higgs Pair Production
  at N$^3$LO}},  {\em Phys. Rev. D} {\bf 98} (2018), no.~11 114016,
  [\href{http://arxiv.org/abs/1811.07906}{{\tt arXiv:1811.07906}}].

\bibitem{Forte:2013mda}
S.~Forte, A.~Isgrò, and G.~Vita, {\it {Do we need N$^3$LO Parton
  Distributions?}},  {\em Phys.Lett.} {\bf B731} (2014) 136--140,
  [\href{http://arxiv.org/abs/1312.6688}{{\tt arXiv:1312.6688}}].

\bibitem{Buckley:2014ana}
A.~Buckley, J.~Ferrando, S.~Lloyd, K.~Nordström, B.~Page, et~al., {\it
  {LHAPDF6: parton density access in the LHC precision era}},  {\em
  Eur.Phys.J.} {\bf C75} (2015) 132,
  [\href{http://arxiv.org/abs/1412.7420}{{\tt arXiv:1412.7420}}].

\end{thebibliography}
\end{document}